Carmine Collina, Giulia Marciani, Ivan Martini, Carlo Donadio, Leopoldo Repola, Eugenio Bortolini, Simona Arrighi, Federica Badino, Carla Figus, Federico Lugli, Gregorio Oxilia, Matteo Romandini, Sara Silvestrini, Marcello Piperno, Stefano Benazzi

Refining the Uluzzian through a new lithic assemblage from Roccia San Sebastiano (Mondragone, southern Italy).

# Refining the Uluzzian through a new lithic assemblage from Roccia San Sebastiano (Mondragone, southern Italy).

Carmine Collina[a1], Giulia Marciani[b,c1*], Ivan Martini[d], Carlo Donadio[e], Leopoldo Repola[f], Eugenio Bortolini[b], Simona Arrighi[b,c], Federica Badino[b,g], Carla Figus[b], Federico Lugli[b], Gregorio Oxilia[b], Matteo Romandini[b], Sara Silvestrini[b], Marcello Piperno[a], Stefano Benazzi[b]

1These authors equally contributed to the paper
*Corresponding author

a. Museo Civico Archeologico Biagio Greco, Mondragone, Caserta, Italy. Carmine Collina: carcollina72@gmail.com; Marcello Piperno: m.piperno@tin.it

b. Università di Bologna, Dipartimento di Beni Culturali. Via degli Ariani 1, 48121 Ravenna, Italy.

Giulia Marciani: giulia.marciani@unibo.it; Simona Arrighi: simona.arrighi@unibo.it; Federica Badino: federica.badino@unibo.it; Eugenio Bortolini: eugenio.bortolini2@unibo.it; Carla Figus: carla.figus3@unibo.it; Federico Lugli: federico.lugli6@unibo.it; Gregorio Oxilia: gregorio.oxilia3@unibo.it; Matteo Romandini: matteo.romandini@unibo.it; Sara Silvestrini: sara.silvestrini6@unibo.it; Stefano Benazzi: stefano.benazzi@unibo.it

c. U. R. Preistoria e Antropologia. Dipartimento di Scienze Fisiche, della Terra e dell'Ambiente, Università di Siena. Via Laterina 8, 53100, Siena, Italy.

d. Dipartimento di Scienze Fisiche, della Terra e dell'Ambiente, Università di Siena. Via Laterina 8, 53100, Siena, Italy. ivan.martini@unisi.it

e. Dipartimento di Scienze della Terra, dell'Ambiente e delle Risorse, University of Naples Federico II, Via Cintia 21, Campus of Monte Sant'Angelo, 80126, Napoli, Italy. Carlo Donadio: carlo.donadio@unina.it

f. University of Naples Suor Orsola Benincasa, Via Suor Orsola, 10 - 80135, Italy. Leopoldo Repola: repolaleopoldo@gmail.com

g. C.N.R, Istituto di Geologia Ambientale e Geoingegneria, 20126, Milano, Italy

**Abstract:**

Roccia San Sebastiano is a tectonic-karstic cave located at the foot of the southern slope of Mt. Massico, in the territory of Mondragone (Caserta) in Campania (southern Italy). Systematic excavation has been carried out since 2001, leading to the partial exploration of an important Pleistocene deposit, extraordinarily rich in lithic and faunal remains. The aim of this paper is to (1) present the stratigraphic sequence of Roccia San Sebastiano, and (2) technologically describe the lithic materials of squares F14 t18, t19, t20; E16 t16, t17, t18 recently recognised as Uluzzian. The stratigraphic sequence is more than 3 metres thick and dates from the Middle to the Upper Palaeolithic. It contains different techno-complexes: Gravettian, Aurignacian, Uluzzian and Mousterian. In the Uluzzian lithic assemblage mostly local pebbles of chert were used in order to produce small-sized objects. The concept of debitage mainly deals with unidirectional debitage with absent or fairly accurate management of the convexities and angles; the striking platforms are usually natural or made by one stroke. It is attested the use of both direct freehand percussion and bipolar technique on anvil in the same reduction sequence. Amongst the retouched tools the presence of two lunates is of note. This study of the Roccia San Sebastiano Uluzzian lithic complexes is significant for understanding the dynamics of the transition from Middle to Upper Palaeolithic in the Tyrrhenian margin of southern Italy.

# 1 Introduction

The Uluzzian is a techno-cultural complex that dates from 45,000 to 40,000 years cal. BP, the period in which Neanderthals were replaced by Modern Humans (MHs). This techno-complex, initially identified and typologically described based on the materials found at Grotta del Cavallo (Salento, Apulia, south eastern Italy) (Palma di Cesnola 1964, 1989, 2004), displays features which are usually thought to be typical of the so-called "modern behaviour", including the presence of colouring substances and the systematic use of bone tools, and ornaments (Arrighi et al., 2020a, 2020b). Moreover, a sharp break has been detected between the Uluzzian and the preceding and partially coeval Mousterian techno-complex both from a technological point of view and in terms of subsistence strategy (Boscato et al., 2011; Boscato and Crezzini, 2012; Marciani et al., 2020, Romandini et al., 2020, Sano et al., 2019). Currently, the Uluzzian is considered a product of MHs

(Benazzi et al., 2011) mainly due to the attribution to *Homo sapiens* of two deciduous teeth found at Grotta del Cavallo in association with the Uluzzian layers (cf. Moroni et al., 2013, 2018; Marciani et al., 2020; see contra Zilhão et al., 2015).

Technologically, the Uluzzian is characterised by the application of "simple" debitage where the striking platform is made by a single or few removals or just a natural or cortical platform is used, and the debitage surfaces are roughly managed. Knapping strategies are dominated by the bipolar technique and mainly geared towards the production of small blades/bladelets and small flakes/flakelets (Ronchitelli et al., 2009, 2018; Riel-Salvatore, 2009; 2010; De Stefani et al., 2012; Moroni et al., 2013; 2018; Villa et al., 2018; Peresani et al., 2019; Marciani et al., 2020). Curved backed implements known as *lunates* or *crescents*, which have been identified as part of projectile-based weaponry (Sano et al., 2019), are the hallmark of this techno-complex (Palma di Cesnola 1964, 2004). Retouched tools mostly consist of a systematic production of end-scrapers (Gambassini, 1997; Palma di Cesnola, 1993).

There are currently 11 sites with a well-dated stratigraphy that attest the presence and the distribution of the Uluzzian along the Italian Peninsula. A cluster of sites is known in Apulia (Cavallo; Uluzzo C, Uluzzo; Serra Cicora; Bernardini), another cluster is documented on the Tyrrhenian side (Cala; Castelcivita; Colle Rotondo; La Fabbrica), and a third group is located in north eastern Italy (Riparo del Broion and Fumane) (Fig. 1).

The discovery of a new Uluzzian site and the detailed study of its archaeological evidence are of paramount importance in contributing to the international debate regarding the technical definition of the Uluzzian techno-complex. This study is valuable in order to shed new light on the relationships between the lithic industries from the above-mentioned sites and their authorship, as well as the role played by the diffusion of the Uluzzian makers in a land previously inhabited by the Mousterian makers. The principal aim of this paper is to describe the site at Roccia San Sebastiano (Lavino et al., 2003; Belluomini et al., 2002, 2007; Piperno, 2006; Collina et al., 2008; Ruiu et al., 2012) and to offer an accurate description of its archaeological sequence, including the technological study of the Uluzzian lithic materials uncovered at the site.

## 2 The Site

### 2.1 Geologic and Geomorphological outline

The Roccia San Sebastiano cave (Figs. 2 and 3) is located in the Municipality of Mondragone (Caserta, northwestern Campania). It is a tectonic-karstic cave which opens at the foot of the southern slope of Mt. Massico (Figs. 2 and 3). The cave opens into Cretaceous limestone rocks attributable to the paleo-geographic units of Mt. Matese - Mt. Maggiore and that form the main part of Mt. Massico. The Mt. Matese - Mt. Maggiore alignment delimits the northern sector of a tectonic depression known as the Campania Plain graben (Aiello et al., 2018).

The surrounding landscape of Mt. Massico is characterised by two plains that originate since the Late Quaternary (Fig. 2): the first one (to the NW) is called the Garigliano River alluvial plain, whilst the second one (to the SE) is the Campanian Plain and includes the Volturno River mouth. These zones are enclosed by the high-morphostructural carbonate relief of Mt. Massico (Billi et al., 1997; Aiello et al., 2018) (Fig. 2). Alluvial plain deposits that characterise these plains consist of Pleistocene-Holocene reworked pyroclastic and fluvial-marine sediments (Aiello et al., 2018).

The present-day morphological features of the overall area are markedly influenced by the violent Campanian Ignimbrite eruption ($^{40}Ar/^{39}Ar$ age: 39.85 ± 0.14 ka; Giaccio et al., 2017) of the Phlegraean Fields. The event generated a vast marine gulf that opened to the northwest, and that was gradually filled by river and marine sediments as well as volcanic debris that came from the erosion of the surrounding hills (Pennetta et al., 2016; Aiello et al., 2018).

## 2.2 The Roccia San Sebastiano cave

The entrance to the Roccia San Sebastiano cave is located about 40 metres asl, near an abandoned limestone quarry (Fig. 3A-C). The cave entrance consists of a narrow passage that widens into a chamber extending in NE direction (Fig. 3B). At the time of the discovery the cave entrance was obstructed by ancient landslide deposits and by residue rocks from the nearby quarry.

The stratigraphic study of the cave filling deposits highlighted the alternation of different phases of deposition in which siliciclastic sedimentation (dominated by collapses and debris accumulations) alternated with phases of carbonate precipitation. In detail, carbonate deposits, in the form of thick flowstones, interbedded with siliciclastic sediments are observed in the innermost areas of the cave (Belluomini et al., 2007). Speleothems, mainly stalactites, are present along the main fractures of the ceiling.

The cave is divided into two distinct parts: i- the shelter (about 12 metres in length and 3 metres in depth. Fig. 4A), and ii- the cave (whose dimensions have not yet been ascertained due to the fact that it is still partially obstructed by reworked sediments, Fig. 4B-C). The shelter was filled with a thick succession of fine- and coarse-grained deposits (allochthonous sediments) and large debris (i.e. autochthonous sediments deriving from the collapse of the ceiling of the cave).

The excavations highlighted a thick sedimentary sequence that can be broadly subdivided into two parts. The stratigraphically upper one (Fig.4A) is composed of coarse-grained sediments and debris, containing a mixture of Palaeolithic and historical artefacts. This evidence, combined with the scarce sediment organisation, indicates that this part of the sequence derives from the accumulation of reworked sediments. These reworked deposits were removed during the first excavations (2001-2010). A carbonate level separates the upper and the lower part of the deposit. The lower part (Fig.4A), consisting of reddish sandy sediments, is characterised by the archaeological deposits *in situ*, rich in faunal and lithics remains. These *in situ* deposits are those investigated in this work and they are subdivided into three main units (see cf. 2.4 for a better description).

**2.3 Research history**

The Roccia San Sebastiano cave was discovered in December 1999 during systematic surveys carried out by C. Collina and M. Piperno as part of a project promoted by the Prehistory Chair of the University of Naples 'Federico II', and the Museo Civico Biagio Greco of Mondragone together with the Soprintendenza Archeologia Belle Arti e Paesaggio di Salerno e Avellino, Benevento e Caserta. In the following years, the cave was included in a geoarchaeological research project aimed at producing a detailed territorial study regarding the dynamics of the frequentation of the area during prehistoric and protohistoric periods (Aiello et al., 2018).

From 2001 to 2010 the systematic excavation (led by MP) was mainly focused on removing the sediment outside the cave and the reworked sediments inside the cave, (which contained a mixture of Upper Palaeolithic, Roman and Medieval artefacts.) The first *in situ* level began to be excavated in 2003 in an area of 6 m$^2$ (F-E 10, 11, 12). It is a Gravettian level, known as C (Collina and Gallotti, 2007; Collina et al., 2008). Furthermore, a test trench 2x1m in E14-E15 was dug in order to understand the potential of the cultural sequence of the cave. The trench is 2.8 m deep and stops where it reaches a sterile layer. It provides the first archeo-stratigraphic description and a reference for the archaeological deposit of the cave (Fig. 5, Tab. 1).

From 2011 to 2019 the excavations (led by CC, Fig. 5) were focused on broadening the E14-E15 trench in F14 in order to evaluate the stratigraphic relation between a large collapsed block (see Fig. 4A) and the archaeological deposit. The excavations were performed in order to understand if the large block was present or not during the prehistoric settlements. The trench in F14 reached the same depth as the trench E14-E15 (2.8 m). The trench was subsequently enlarged an extra 30 cm$^2$ in E16 (Fig. 5).

The lithic materials of Roccia San Sebastiano were initially studied in order to assess the potential of the deposit and were then published in a local journal (Collina and Piperno, 2011). A specific focus was given to the Gravettian level C (Collina, Gallotti 2007; Collina et al., 2008). The presence of a Uluzzian component at Roccia San Sebastiano was identified in 2015 and only recently was a preliminary study presented at the UISPP congress in Paris in 2018 (Collina and Piperno, 2018a) and in local journals (Collina and Piperno, 2018b; Collina et al., 2020). Subsequently, the technological study was performed under the egis of the ERC project SUCCESS "The earliest migration of Homo sapiens in southern Europe".

### 2.4 Stratigraphic and archaeological sequence

The archaeological sequence excavated at Roccia San Sebastiano (found in trench E14-15) can be divided into three main units (labelled Unit 1 to 3) based on their overall stratigraphic features and on the archaeological materials that were discovered. Each unit is in turn subdivided into archaeological/lithostratigraphic sub-units that are shown in Figure 6D and briefly described below in inverse stratigraphic order:

- Unit 1: the unit is mainly made up of brownish sandy silt deposits with occasional carbonate concretions. The uppermost portion of the unit displays a coarser matrix, while, traces of charcoal levels and carbonate cobbles occur at the base of the unit. Unit 1 consists of the sub-units C-Ca (recent Gravettian), Cb (Gravettian with Noailles burins), and Cc (Early Gravettian);
- Unit 2: the unit consists of reddish-brown sandy silt deposits with the occasional presence of limestone pebbles. The sediments slope slightly towards the interior part of the cave. Unit 2 includes the sub-units Cd (Initial Gravettian), and Ce (Aurignacian with Dufour bladelets). A large block of collapsed limestone separates the sub-units Ce and Cd;
- Unit 3: the unit consists mainly of dark sandy, silty deposits, rich in organic matter and with abundant remains of fauna. Unit 3 is composed of two sub-units, labelled Cf (Uluzzian) and Cg (Final Mousterian). In greater detail; the sub-unit Cf is made up of two lithostratigraphic strata; the upper is made of reddish compact clay with scarce limestone debris, whilst the lower is mainly composed of yellowish clay.

During the excavations each sub-unit was dug into artificial spits of 5-7 cm, which followed the slope of the layers, the presence of archaeological features and/or the location of archaeological materials. These spits were indicated by "t" plus a consecutive number (t1, t2, t3…).

In trench E14-E15 seven major cultural phases were identified, their main features are laid out in Table 1 and illustrated in Fig. 6A-B-C-D. The table also includes the relationship between spits dug in the two new trenches F14 and E16 (excavation CC) and those previously dug in the old reference trench E14-E15 (excavation MP).

Aiello et al., (2018) report two calibrated chronological dates for the stratigraphic sequence in E14-E15 (Tab. 2). The oldest date (R_Date Rome-2111) is 43,680 - 42,190 (68.2%) cal BP– 44,740 – 41,700 (95.4%) cal BP based on a bone fragment collected in sub-unit Cg-t29-34 (Final Mousterian). The second date (R_Date Rome-2447) is 23,870 - 23,320 (68.2%) cal BP – 24,070 – 23,020 (95.4%) cal BP based on a bone collected in sub-unit Ca-t1-4 (Recent Gravettian). Dates were calibrated through OxCal 4.3 using Intcal13 data (Reimer et al., 2013).

## 3. Material and Methods

### 3.1 Material

This study is focused on the technological analysis of the lithic material uncovered in spits t18, t19 and t20 trench F14, and spits t16, t17, and t18 trench E16 (Tab. 3). Twenty-two items were excluded from the study because they clearly pertained to an upper Gravettian level, this intrusive presence could be related to post-depositional processes.

### 3.2 Lithic technology

The lithic assemblage was analysed using the technological approach which is critical in order to place each object in a precise technical context (Boëda, 1991, 2013; Geneste, 1991; Inizan et al., 1999). This approach is used to identify all the technical and economic processes performed during the production of a tool: from the acquisition of raw material, through the different stages of the manufacturing process, to the use and subsequent disposal of the tool.

The categorical variables chosen to perform the study, based on macroscopic evidence, were: lithotypes (chert, cherty limestone, limestone, quartzite, quartz, sandstone), granulometry (fine, coarse), raw material geological origin (pebbles, slabs, nodule), raw material colour (the colour of cortex and the colour of the inner portion of the items), type of patina, presence of combustion traces (yes, no).

From a morphometric point of view, all items were divided into five dimensional classes (DC) (first; 1-50 $mm^2$, second; 50-100 $mm^2$; third: 100-150 $mm^2$; fourth: 150-200 $mm^2$; fifth: > 200 $mm^2$) based on the area covered by each specimen size (Marciani et al., 2016; Spagnolo et al.,

2016). Length, breadth, thickness of items with an area larger than 50 mm$^2$, were also measured according to their technological axis. When this was not possible, the longest measurement was conventionally regarded as the length. The presence and location of macro-traces were assessed both by the naked eye, and by the use of a magnifying glass. All the artefacts were assigned to integrity classes based on the location of identified fractures (integer, composite, distal, lateral, mesial, proximal), as well as to technological classes: core, flake, micro-flake (integral flakes of the 1-2 DC), debris (fragmented pieces, altered pieces, un-orientable pieces), hammer-stone.

Flakes were grouped into cortical flakes (100% cortical) and semi-cortical flakes (between 50-75% of cortex coverage). Flakes presenting less than 50% of cortex coverage were grouped into flakes, long flakes, and blades based on the ratio between length and breadth (ratio between 0 and 1.5 = flake; ratio between 1.5 and 2 = long flake; greater than 2 = blade) (Laplace, 1966). Evidence of a specific technological category of flakes called "*bâtonnets*" (Tixier, 1963) was found, which is a thick flake/blade obtained by fragmenting a core through bipolar technique. For each flake, the orientation of dorsal scars and the quantity of cortex; the profile section; the type of butt and bulb; the position of the impact point; the presence of abrasion and parasital scars, were registered. Attention was paid to the proximal portion of the flake in order to evaluate the percussion technique. All the above mentioned technical traits were used to identify the concept of debitage (Boëda, 2013).

For each core, we recorded the geological origin (pebble, block, slab) and the morphology of the raw block; the volumetric conception used for core exploitation (Boëda, 2013); the number of exploited faces; the hierarchy of surfaces (yes, no); the number, type, location and mode of preparation of the striking platform; and, lastly; number, direction, and chronology of the scars on the debitage surface (Inizan et al., 1999). All the photos of the lithic materials were taken by a Fujifilm XT3 with a macro lens 80xx, using a portable RTI Dome (a device developed by Tomasz Łojewski, AGH University of Science and Technology). The graphic elaboration was made on Corel Draw Graphic Suite X7.

## 4 Results

### 4.1 Raw Material

The lithic materials display fresh margins (only 22 items have worn edges). About 60% of the 1052 items assigned to Dimensional Class 3 or above, present a white patina. It must be underlined that the presence of 8 items with a double patina indicates the reuse of older materials. The presence of pieces showing traces of combustion (shining surfaces, de-silicified portions, or presence of fire-hole marks) is considerable, especially in square E16, in spit t17 (Tab. 4).

The vast majority of the assemblage (90.1%), considering all spits at once, is made out of chert mainly of three types; 1. fine-grained, white/azure, opaque chert; 2 fine-grained, glossy beige or grey chert; 3 fine grained, opaque, red/orange chert. This is followed by a very low percentage of limestone (5%) and cherty limestone (i.e. a limestone containing a high percentage of silica, 3.9%), with a sporadic presence of radiolarite (0.6%) and quartz-arenite (0.3%) (Tab. 5). There is evidence suggesting a preference for fine-grained raw materials (83.5 % of the total; (Tab. 6, Fig. 7). The fine-grained, white/azure, opaque chert (Fig. 7A-B) was collected locally, whilst the fine-grained, glossy beige or grey chert (Fig. 7D-E) closely resembles pebbles of Apennine origin (Aiello et al., 2018). Fine-grained, opaque, red/orange chert (Fig. 7F) could possibly come from the Scaglia Rossa Formation Umbro Marchigiana (Aiello et al., 2018). In most cases, the type of raw material is small-sized pebbles (3–5 cm) that are easy to find even nowadays along the sandy and pebbly shores of some rivers located near the site, i.e., along the plains facing Mt. Petrino and Mt. Massico, and also about 30 km from the site in the area of Triflisco on the Volturno River (Aiello et al., 2018; Vitale et al., 2019).

## 4.2 Reduction sequence

More than half (56.4%) of the assemblage is fragmented (Tab. 7), and mostly consists of items belonging to the smaller dimensional classes. Nevertheless, classes comprising larger pieces also present cases with a high degree of fragmentation, which is possibly ascribable to the nature of the raw material (e.g. fractured inner layers) or to the chosen percussion technique. (i.e. a bipolar technique on anvil which produces a high number of fragmented items).

The first two dimensional classes (which correspond to the majority of debris and micro-flakes, i.e. the waste of debitage), are the most represented ones across all spits (first DC: 1-50 mm$^2$ = 27.3%, second DC: 50-100 mm$^2$ = 28.5), followed by the bigger items pertaining to the fifth DC (> 200 mm$^2$ = 18.8) (Tab. 8). The high quantity of debris and micro-flakes (namely items of the first DC: 1-50 mm2, second DC: 50-100 mm2) suggests an intensive activity of *in situ* debitage, that is confirmed by the presence of a great number of flakes and cores (47) (Tab. 9). A hammer-stone has also been found which exhibits traces of percussion on one side, and, on the other, evidence of its use as a core to obtain elongated flakes (the chronology of these two actions cannot be defined). A specific technological analysis was performed on the flakes and cores in order to obtain detailed insights into reduction sequences, the management of core reduction, the objectives of debitage, and on flaking technique.

### 4.2.1 Cores

The natural blocks of raw material that were chosen to be flaked, were of varying shapes, dimension (Fig. 8) and each block belonged to one of three main types: flakes (Fig. 9C), broken pebbles (Fig. 9A), or fragments (Fig. 9B) (Tab. 10). The dimension of raw mass is quite small (Fig. 8), and the reduction sequences are short (Tab. 11), i.e. from the selected block only 2 or 3 flakes were extracted, and without any management of the lateral and distal convexities. The striking platform was made, either using a single stroke or just a few strokes (Fig. 9B-C), or, using an unprepared or cortical platform (Fig. 9A). 35 cases display the use of a single striking platform, whilst in 10 cases two opposing striking platforms were used. There are only 2 occurrences of more than 2 striking platforms. The cores were discarded at an intermediate (25 items) or final stage of reduction (20 items). The vast majority of knapping series is unidirectional (39 items). It is worthy of note that, within the same reduction sequence, both direct freehand percussion and bipolar percussion on anvil are present.

### 4.2.2 Bipolar technique of debitage

The bipolar percussion technique on anvil generates products characterised by a number of specific traits. For example, a rectilinear longitudinal profile of the ventral face; similar ventral and dorsal faces; pronounced ripple marks; shattered point-form or linear butts; diffused impact points; sheared bulbs of percussion; and the presence of a parasitical scar (Fig. 10) (i.e., Barham, 1987; Knight, 1991; Guyodo and Marchand, 2005; Grimaldi et al., 2007; Bietti et al., 2010; Bradbury, 2010; Soriano et al., 2010; Moroni et al., 2018; Horta et al., 2019; Marciani et al., 2020). The materials from Roccia San Sebastiano, characterised by clear signs of this percussion technique, consist of 362 items, which can be grouped into several technological classes (Tab. 11, Fig. 11) (cores, flakes, micro-flakes and debris). The typical products resulting from bipolar reduction were identified based on the above-mentioned specific traits (Barham, 1987; Knight, 1991; Guyodo and Marchand, 2005; Grimaldi et al., 2007; Bietti et al., 2010; Bradbury, 2010; Soriano et al., 2010; Moroni et al., 2018; Horta et al., 2019; Marciani et al., 2020). These products are either thin, small, straight items, or thick items with quadrangular cross-sections (*bâtonnets*). Traits on the proximal portion of the flakes (Fig. 10, Tab. 12) are of particular importance in defining the percussion technique.

### 4.2.3 Debitage products

The result of this straightforward, yet only partially predetermined reduction sequence, is a considerable variety in the objects' size, morphology and edge delineation. Debitage products (Fig. 12) include both cortical and semi-cortical flakes, attesting to the beginning of reduction sequences, as well as a large amount of flakes, long flakes and blades (Tab. 13). Many items are fragmented, and items with composite fractures are the most represented category followed by proximal and distal items (Tab. 13). This fragmentation degree could be explained by the raw materials characteristics and by the use of the bipolar technique, but also to the chosen debitage modality. The latter modality is simple, without careful management of striking platforms and debitage surfaces, hence increasing the probability of encountering knapping mistakes (plunging or hinged accidents) and broken flakes. Most flakes exhibit a rectilinear profile (76.5%), but some specimens also have a wavy (9.7%), convex (8.3%), concave (3.8%) or twisted (1.7%) profile. Most flakes (more than half of the sample; Tab. 14) display unidirectional scars and flat, linear or cortical butts (Tab. 15), all of which are fully compatible with cortical un-prepared striking platforms and the core opened by the use of a single-stroke.

The present assemblage contains 122 retouched pieces of which 63.3% are flakes, 33.6% are debris, and 3.3% are cores (Tab. 16). Side scrapers (23.3%) and splintered pieces (19.7%) are the main types present. The presence of two lunates (Fig. 13) is remarkable. They both possess a curved backed side faced by a rectilinear cutting edge, but they are very different in size. Side-scrapers and end-scrapers were mainly made from flakes, as were the two lunates. In some cases, it was not possible to evaluate the original retouched blank.

## 5 Discussion

### 5.1 Interpreting splintered pieces

One of the most challenging problems, when analysing products of the bipolar technique, is understanding the meaning of the term 'splintered pieces' (also referred to as 'scaled pieces', equivalent to the French *pièces esquillée;* Brézillon, 1983). 'Splintered pieces' are typologically described in scientific writing (i.e. Brézillon, 1983, Inizan et al., 1999, Le Brun-Ricalens, 2006), as quadrangular, irregular items which are frequently splintered, sometimes bifacially. Such splintered marks, generally, appear on the two opposite extremities of the piece, whilst, more rarely, they can be seen on either just one extremity, or on all four extremities. The term 'splintered pieces' has come to cover a variety of different objects: both the above mentioned typologically described items and all the splintered fragments or flakes produced by the use of the bipolar technique (Tixier, 1963; Inizan et al., 1999). Moreover, the role of these typologically defined 'splintered pieces' (*sensu* Brézillon, 1983) remains open to interpretation: should they be considered as cores or as tools? (i.e. Inizan et al., 1999, Le Brun-Ricalens, 2006, Villa et al., 2018, Horta et al., 2019).

When the role of the splintered pieces is the core it means that the scaled features are scars that derive from the purposeful extraction of flakes, to reiterate – these flakes are the aim of the debitage. The splintered pieces/cores are generally thick, with flake scars extending over the full length of the core. Moreover, the striking platform is rarely preserved intact, and one or both ends display battering with step fractures (Le Brun-Ricalens, 2006; Villa et al., 2018). The flakes obtained by this kind of production are generally small (< 3 cm), with the proximal and distal ends possibly smashed, and they show small step scars and splintering close to the impact area and/or at the distal end (Le Brun-Ricalens, 2006; Villa et al., 2018).

When the role of the splintered pieces is the tool, there are two variants: 1. a tool that has been intentionally retouched or 2. a tool that has been 'unintentionally' retouched. In the first case the retouching of the tool was a predetermined and intentional action which aimed to thin the edge (Ranaldo et al., 2017). In the latter case the retouching came about due to the use of the tool. Namely, a tool whose "retouched" edge is the *a posteriori* result of an unintentional activity, i.e. due to the use of the tool as an intermediary/wedge (Villa et al., 2018). According to this second hypothesis the splintered piece (hafted or not) can be used as an intermediary "tool" in order to perform a variety of activities such as fracturing, splitting, dividing, and cutting through the use of bipolar or direct percussion (le Brun-Ricalens, 2006; Langejans, 2012). Thus, they present traces created by violent percussion and also have different morphologies which are created by more or less prolonged use (Le Brun-Ricalens, 2006). Examples of hafted tools, that resemble splintered pieces and are used for specialised activities, are the Amazonian *dentes de rallador* (Duarte-Talim, 2012; 2015) and the insets of the Neolithic *tribulum* (Le Brun-Ricalens, 2006).

The obtained splintered pieces and fragments acquire different roles according to the scenario: core vs tool. If the splintered piece (*sensu* Brézillon, 1983) is a core, the obtained splintered flakes and fragments become the aim of the debitage. If the splintered piece (*sensu* Brézillon, 1983) is a tool, the obtained splintered flakes and fragments become, simply, waste. Moreover, as well as these difficulties in identifying the role of the splintered piece, it is important to note that there is also a certain fluidity between the various roles. This debate cannot be resolved without an integrated approach that encompasses technological, techno-functional and traceological analyses. The entire context of the lithic assemblages needs to be taken into account, and an approach that focuses on the intentions of prehistoric craftsmen is needed.

Based on the study of the materials at Roccia San Sebastiano, the presence of several technological classes (cores, flakes, micro-flakes, debris) all showing features of bipolar flaking technique lead us to lean towards a reading of an intentional and deliberate use of this technique in the reduction sequence. Therefore, the splintered pieces present at Roccia San Sebastiano may be considered cores. Only few cases portray regular small scars that usually affect a single edge. At the present date it cannot be said exactly which use created these scars. A programmatic use-wear campaign must be performed.

## 5.2 Attribution of the lithic assemblage at Roccia San Sebastiano to the Uluzzian techno-complex

The lithics from the studied levels at Roccia San Sebastiano fall within the range of the Uluzzian techno-complex (Moroni et al., 2018; Marciani et al., 2020) (Tab. 17). At Roccia San Sebastiano the preferred source of raw material is the local pebbles of chert that can be found near the site. The reduction sequence is characterised by unidirectional debitage, and the striking platform is either cortical or made with one or a few strokes. Distal and lateral convexities of the debitage surface are rarely managed. Any kind of raw block displaying angles and guide ribs – which make it suitable for flaking – is selected and used (Fig. 14). This kind of debitage is quite simple because it takes advantage of the technical characteristics of the raw blocks, and only makes use of a portion of the block itself (additional debitage according to Boëda, 2013). The integrated (*sensu* Boëda, 2013) production concepts (i.e., Levallois and discoid typical of the Mousterian or laminar lamellar reduction systems typical of the UP) are missing. The debitage at Roccia San Sebastiano is aimed at the production of a variety of flake shapes and elongated pieces with a generally low degree of standardisation. Bipolar knapping on anvil is the most frequent percussion technique, combined with unipolar direct freehand percussion. Furthermore, analysis suggests that bipolar and direct percussion are part of the same unidirectional reduction sequence. Retouched tools are mostly side-scrapers, end-scrapers, as well as lunates (although just two of them were found) (Fig. 14).

The Uluzzian sites in Italy, which until now have been studied with a more or less specialised technological approach are: Cavallo level EIII (Moroni et al., 2018), Uluzzo C level 3, 15, 17 (unpublished data), Castelcivita levels rsa'', rsi, rpi, pie (Gambassini, 1997; unpublished data), Roccia San Sebastiano levels F14 t18, t19, t20; E16 t16, t17, t18), Colle Rotondo (Villa et al.,

2018), Fabbrica 2 (Villa et al., 2018), Broion levels 1g, 1f (Peresani et al., 2019), Fumane level A3 (Peresani et al., 2016; 2019). These collections show some internal differences in the mode of production, possibly due to the different chronological phases that they come from, or to different local adaptation. However, several common features can be underlined (Tab. 17). More specifically, the use of local raw materials and the dominant use of the bipolar technique on anvil are two distinctive features of the Uluzzian that can be noted in all the sites taken into consideration (Cavallo, Uluzzo C, Roccia San Sebastiano, Castelcivita, Colle Rotondo, Fabbrica, Broion, Fumane). The presence of a bipolar technique and direct percussion is clear. However, it is necessary to further explore these two components, to see whether they are part of the same sequence (in the case of Roccia San Sebastiano) or rather represent two distinct reduction sequences. Common in all the Uluzzian sites is the little managed concept of debitage such as unidirectional and orthogonal methods. The integrated concepts such as the Levallois and discoid debitage which characterised the Mousterian are missing in the Uluzzian. The integrated, very curated debitage such as the blade/bladelets reduction systems typical of the Upper Palaeolithic are almost completely absent in the sites (with the exception of a presence at Broion and Uluzzo C). The concept of debitage at Riparo del Broion is unidirectional (performed by bipolar knapping technique) and there is also present evidence of a lamino-lamellar production (Peresani et al., 2019). Fumane is a case apart, as it is the only Uluzzian site which presents a Levallois component and a low presence of bipolar technique (Peresani et al.,2016; 2019). The assemblage of La Fabbrica is characterised by a unidirectional and orthogonal debitage and the striking platforms are flat. The exploitation was performed on only one or two adjacent debitage surfaces (Villa et al., 2018). At Colle Rotondo, unidirectional, bidirectional debitage are present, and the striking platforms are either cortical or made using one or several removals (Villa et al., 2018). The levels of Castelcivita are characterised by unidirectional and orthogonal debitage which takes advantage of the guide ribs of the blocks (further studies of the levels are ongoing). The reduction sequence at Roccia San Sebastiano is characterised by unidirectional debitage with a cortical or a flat striking platform. Uluzzo C is characterised by a more managed unidirectional volumetric debitage and a low controlled unidirectional debitage, the relation between these two components is still under investigation.

Considering the retouched tools, the Uluzzian is characterised by a systematic production of end-scrapers (Palma di Cesnola 1964; 1989; 2004), the presence of which is noted at Roccia San Sebastiano, Cavallo, Uluzzo C, Castelcivita, Collerotondo, Fabbrica, Broion, Fumane. The lunates are generally considered to be the 'hallmark of the Uluzzian'. This definition was coined after the recurrence of lunates at Cavallo, where the Uluzzian techno-complex was first identified and defined (Palma di Cesnola 1964, 1989, 2004). The lunates are considered the defining feature of the Uluzzian due to their specific construction (a retouched back opposite to a rectilinear cutting edge). And, most importantly, they are considered the hallmark because this tool made its debut in Europe in the Uluzzian, being absent both in the late Mousterian and the Protoaurignacian. However, the number of lunates (not considering the backed items but only considering Gm1- geometric 1, sensu Laplace, 1966) varies widely between the Uluzzian sites: ranging from 0 to 23. Specifically, at Cavallo (Palma di Cesnola's 1963-64 excavations) in layer EIII there are 10 lunates (1.6% of the retouched tools – 640) (Palma di Cesnola, 1966); in layers EII-I there are 23 (9.9 % of the retouched tools - 233) (Palma di Cesnola, 1966); finally, in D there are 5 (2.4% of the retouched tools - 212) (Palma di Cesnola, 1966). There are 4 lunates at Broion f-g (6.6% of the retouched tools - 61) (Peresani et al., 2019); 4 at Colle Rotondo (4.2% of the retouched tools – 95) (Villa et al., 2018); 7 at Castelcivita (0.9 % of the retouched tools - 775) (Gambassini, 1997); 2 at La Fabbrica (1.8 % of the retouched tools – 113) (Villa et al., 2018); 1 at Cala (0.7 % of the retouched tools – 134) (Benini et al., 1997); 2 at Uluzzo C (2.7 % of the retouched tools, levels C and D of the Borzatti's excavation 1964) (Borzatti von Löwenstern, 1965); and 0 at Serra Cicora (0% of the retouched tools – 81) (Spennato, 1981). The lunates display low number also at Roccia San Sebastiano 2 lunates (1.6% of the retouched tools - 122). According to Palma di Cesnola, the presence of lunates significantly decreases by the end of the Uluzzian, this analysis mainly being based on Grotta del Cavallo (Palma di Cesnola, 2004). In a recent paper Sano and colleagues (2019) proved that lunates found at Grotta del Cavallo were armatures of projectile weapons.

**5.3 Key technical concepts defining the Uluzzian techno-complex**

When considering the Italian Uluzzian sites (Cavallo, Uluzzo C, Castelcivita, Colle Rotondo, Fabbrica, Broion, Fumane) with a technological approach, three major characteristics emerge: 1) the conceptualisation of production, 2) the bipolar technique, and 3) the idea of simple production for complex tools.

*1 Conceptualisation of production*

The general idea underlying the shift to Uluzzian industries is the loss of the long lasting technical tradition that preceded it. That is to say, fully predetermined reduction systems disappeared, and were replaced by new technical traditions characterised by simpler production. The previous Mousterian techno-complex (apart from regional and local variants) used concepts of debitage characterised by a high degree of attention towards production, in order to obtain predetermined objectives of debitage. Mousterian assemblages also exhibit a high degree of attention to dimension and the morphology of the objectives, and also exhibit a predetermination of cutting edges. The Levallois concept, indeed, is a clear example of producing fully predetermined items from both a qualitative and quantitative point of view (Boëda, 1994). The Uluzzian is characterised by a simpler production which takes full advantage of the technical qualities exhibited by the selected raw block, while deliberately overlooking any elaborate management of the volume aimed at controlling the technical characteristics of the output. The lack of clear predetermination in designing the objective is balanced by the selection of an appropriate raw block on which a bipolar technique is applied.

*2 Bipolar technique*

The bipolar technique allows one to obtain the target product from any kind of raw block without any previous preparation of the striking platform, and without any management of angles or convexities. Although this technique reduces the predetermination over the morphology of the products, it also allows the toolmaker to obtain a rectilinear profile, absence of prominent percussion bulbs, rectilinear cutting edges, and thin flakes. The unidirectional debitage produced through the bipolar technique would generate several products with just one or a few strikes. The initial choice of the volume to be flaked is the one stage in which predetermination over the volumetric features of the product is expressed (e.g. small pebbles to obtain small flakes; the edge of a flake to obtain bladelets). Size predetermination is entirely bound to the choice of the initial volume. Moreover, according to Le Brun-Ricalens (2006), the consistent recurrence of splintered pieces within prehistoric assemblages has to be attributed both to a lower technical investment and a higher probability of success, in addition to a higher versatility which makes them ideal complementary objects to be combined with other items.

*3 Simple debitage for complex tools*

Despite this production being simple in terms of the concepts and techniques involved, it actually brings considerable technical advantages and lower technical requirements. The technical advantages are related to the higher versatility in terms of initial supports, obtained goals, and debitage products, and the easily produced rectilinear profiles. It is possible that Uluzzian flakes and blades or bladelets were components of composite instruments (Sano et al., 2019; Moroni et al., 2013, 2018; Marciani et al., 2020). The creation of composite tools would have required the ability to manage several domains (i.e. Simondon, 2017, Arthur, 2009): knapping expertise; specific know-how in the extraction and production of adhesives (Boëda et al., 1996; Koller et al., 2001; Wadley et al., 2009; Charrié-Duhaut et al., 2013; Zipkin et al., 2014; Groom et al., 2015; Gaillard et al., 2016; Kozowyk et al., 2016; 2017a; 2017b); building the hafting (Gibson et al., 2004; Rots and

Williamson, 2004; Rots, 2010; Sykes, 2015); control over the *chaîne opératoire* related to other raw materials (e.g. wood, bones) (Gibson et al., 2004; Rots and Williamson, 2004; Rots, 2010; Sano, 2016); and finally the precision required to harmonically assemble all the parts to obtain a functioning tool. The recent study of the lunates at Grotta del Cavallo as parts of composite projectile weapons is therefore very significant (Sano et al., 2019). Consequently, the idea of simple and slightly predetermined production should not be considered as a lack of knowledge, but it should rather be interpreted as the result of a behavioural change, a different and novel way of conceiving tools and technology.

**6 Conclusion**

The debate regarding the technological definition of the Uluzzian techno-complex is becoming more and more focused thanks to current new studies (Moroni et al., 2018; Villa et al., 2018; Peresani et al., 2019; Arrighi et al., 2020a, 2020b; Badino et al., 2020; Marciani et al., 2020; Romandini et al., 2020). Given the scarcity of Uluzzian stratified sites and the importance of this techno-complex for the "transition" phenomenon, the newly discovered presence of Uluzzian lithic materials in the cave at Roccia San Sebastiano is critical in order to understand the dynamics of the Middle to Upper Palaeolithic transition in Italy. Moreover, the site at Roccia San Sebastiano, located between the Castelcivita and the Colle Rotondo prehistoric sites, fills a geographical gap along the Tyrrhenian side of Italy.

More specifically, this research allowed for the presentation of the original technological analysis of the lithic materials at Roccia San Sebastiano within the framework of the stratigraphy of the site. In conclusion, the main technical characteristics of this assemblage are: 1) the use of local pebbles of chert available in sources near the site; 2) the selection of raw blocks presenting angles and guide ribs appropriate for knapping; 3) simple reduction sequences characterised by unidirectional debitage, where the striking platform is cortical or made with one or a few strokes, and the lateral and distal convexities of the debitage surface are rarely managed; 4) the production of a variety of flake shapes and elongated pieces with a generally low degree of standardisation; 5) the use of bipolar and direct percussion in the same reduction sequence; 6) the occurrence of lunates and end-scrapers. These technical features of the lithic materials at Roccia San Sebastiano (trench F14, spits t18, t19 and t20; and trench E16 spits t16, t17, and t18) mean that the site fully ascribes to the Uluzzian techno-complex and permits us to refine our understanding of the Uluzzian technical structure. When comparing Roccia San Sebastiano to the other Italian Uluzzian sites a remarkable technological "cohesion" within this techno-complex can be noted. Notwithstanding the superficial divergences (due to the low standardised shapes and dimensions of cores and debitage products), it is worth noting the number of significant traits that these sites have in common. These are 1) the conceptualisation of production, 2) the deliberate selection of bipolar technique, and 3) the idea of "simple" production for complex tools. This "cohesion" found in lithic technology also emerges from the characteristics of ornaments (Arrighi et al., 2020a, 2020b) which point towards shared attitudes of these groups over a wide geographical area and diverse geomorphological contexts.

**Acknowledgements**

The authors are very grateful to the Municipality of Mondragone for supporting and funding the excavations and would like to thank the Museo Civico Archeologico Biagio Greco, Mondragone, Caserta, Italy which provided funding, logistic support and welcomed the researcher since 1999. An acknowledgement to the Soprintendenza Archeologia Belle Arti e Paesaggio di Salerno e Avellino, Benevento e Caserta. This project has been realised through funding from the European Research Council (ERC) under the European Union's Horizon 2020 research and innovation programme (grant agreement No 724046) http://www.erc-success.eu/. The sedimentological and stratigraphic analysis were funded by the National Geographic Society/Exploration Grant Program (grant NGS-61617R-19 to I. Martini). GM would like to thank M. Santos, L. Carmignani, D. Aureli, M. Rossini, M. Roussel and M. Soressi for the interesting discussions and insights; and S. Ricci, M. Santos for help in photo editing. We also would like to thank Mandala Macgregor for her careful English proof-reading. A final thanks must be made to the two referees whose constructive advice significantly improved the article.

**Author contributions**

S. Benazzi founded the study by the ERC grant agreement No 724046.

Scientific direction of excavation at Roccia San Sebastiano: C. Collina, M. Piperno.

Conceptualisation: C. Collina, G. Marciani.

Original draft: G. Marciani.

Lithic technology: G. Marciani, C. Collina.

Stratigraphic analysis: C. Donadio, C. Collina, I. Martini.

Geomorphological survey: C. Donadio.

Longitudinal section of the cave: L. Repola.

English proof-reading: E. Bortolini.

**Figure Captions**

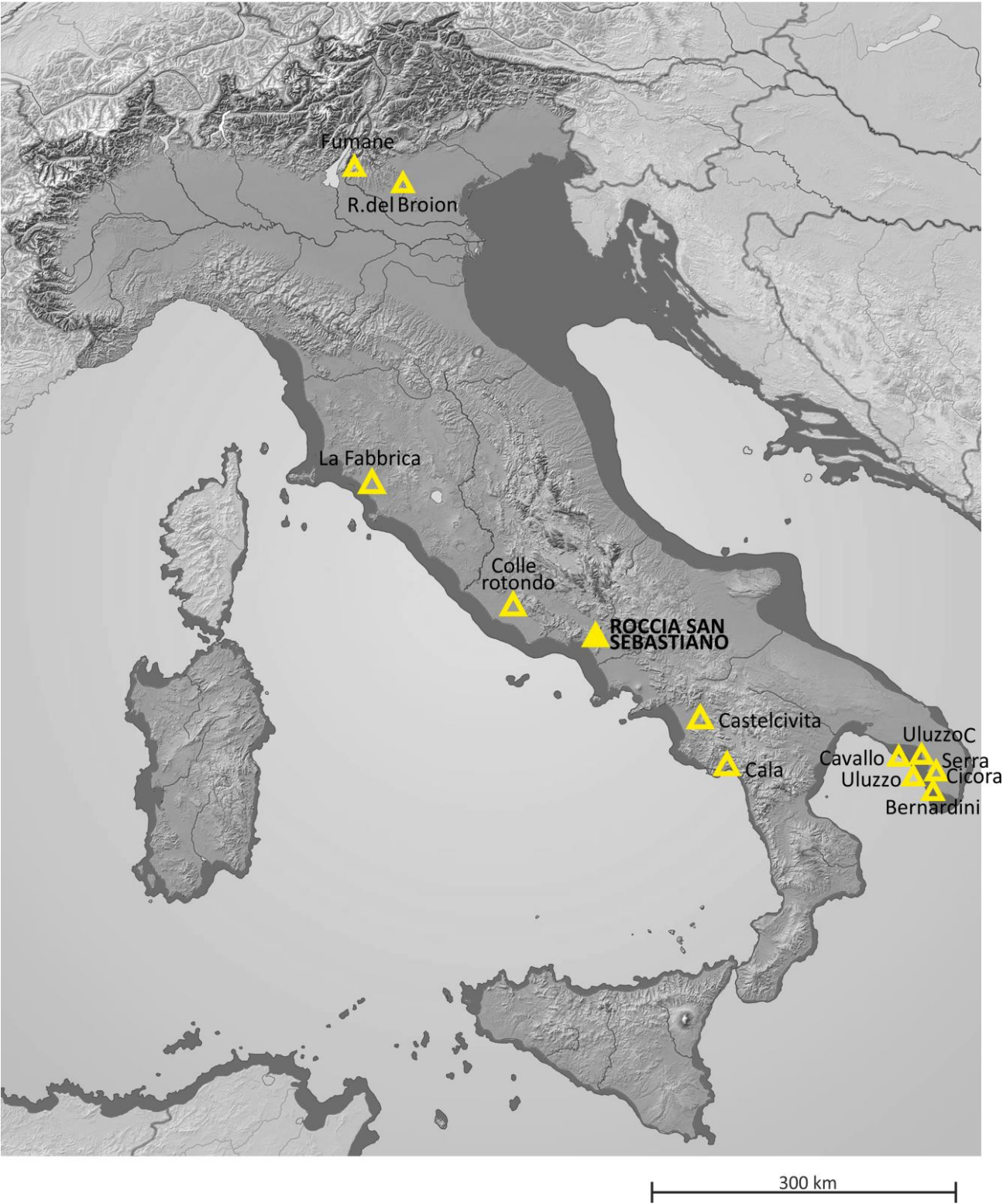

**Figure 1:** Localisation of the Roccia San Sebastiano cave and the other Italian Uluzzian sites in the stratigraphies.

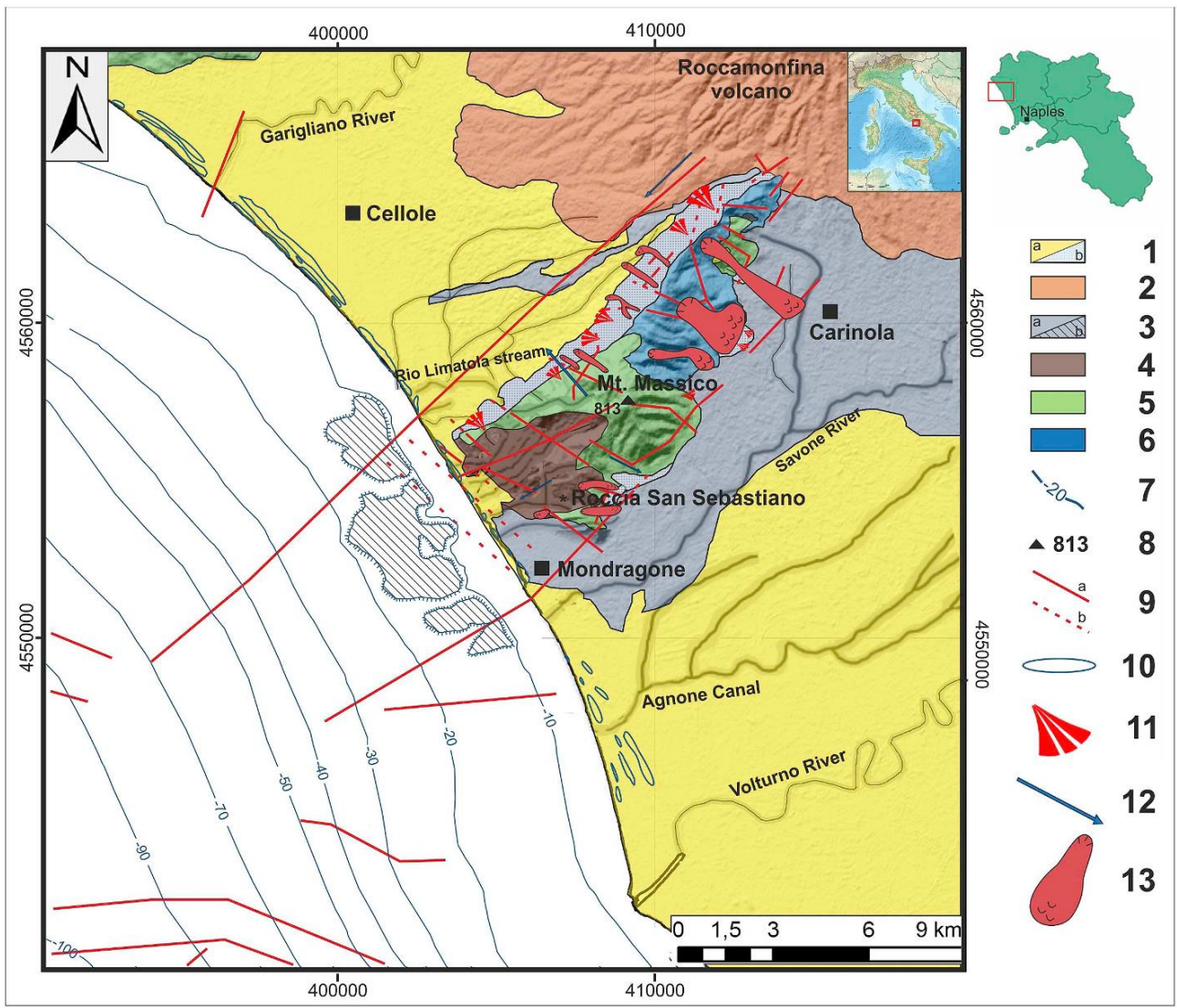

**Figure 2:** Geological map of the coastal plain of Mondragone (from Aiello et al., 2018): 1, sedimentary rocks: a, reworked pyroclastics, fluvial-marine, lacustrine and aeolian deposits of the Campania Plain; b, silty-sandy seafloor of the Gaeta Gulf (Quaternary); 2, lavas and pyroclastics of the Roccamonfina volcano (Middle-Late Pleistocene); 3, Campanian Ignimbrite: a, continental; b, submerged; 4, terrigenous deposits in flysch facies (Miocene); 5, limestone and dolomitic limestone, interbedded with levels of conglomerate in clay matrix (Cretaceous); 6, oolitic limestone and dolomite (Upper Lias); 7, isobath (m bsl); 8, altitude (m asl); 9, fault: a, certain; b, presumed or concealed; 10, dune ridge; 11, alluvial fan; 12, stream incision; 13, landslide pile.

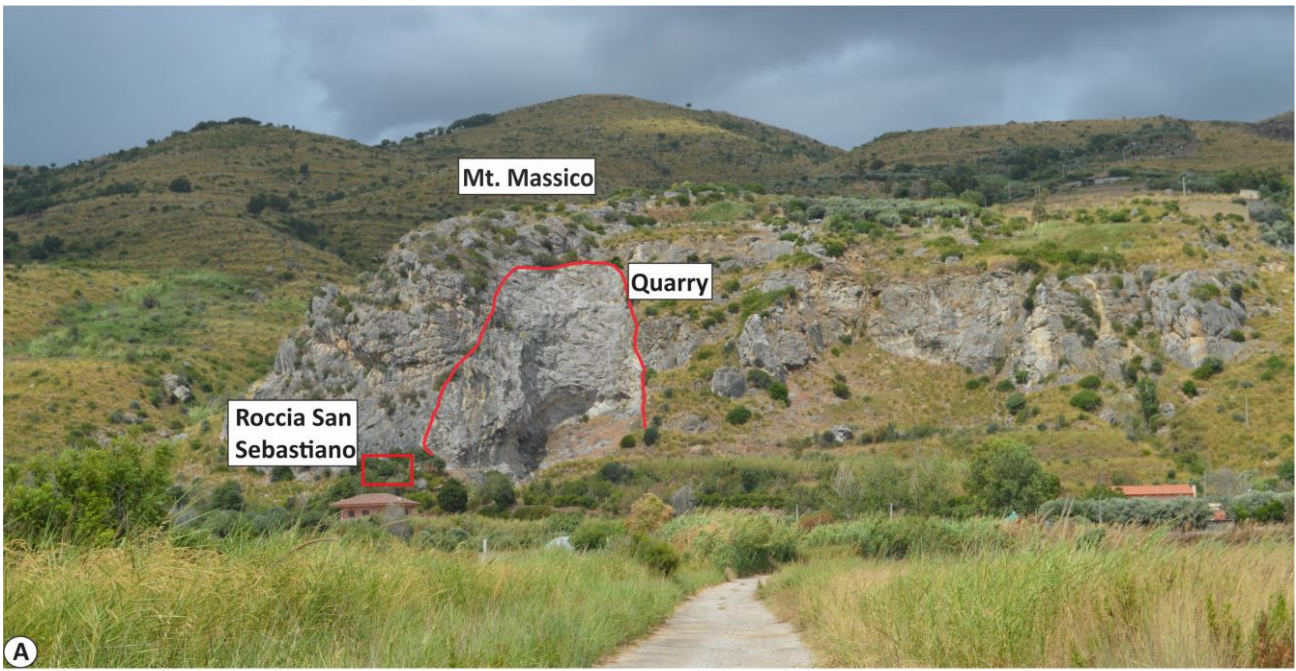
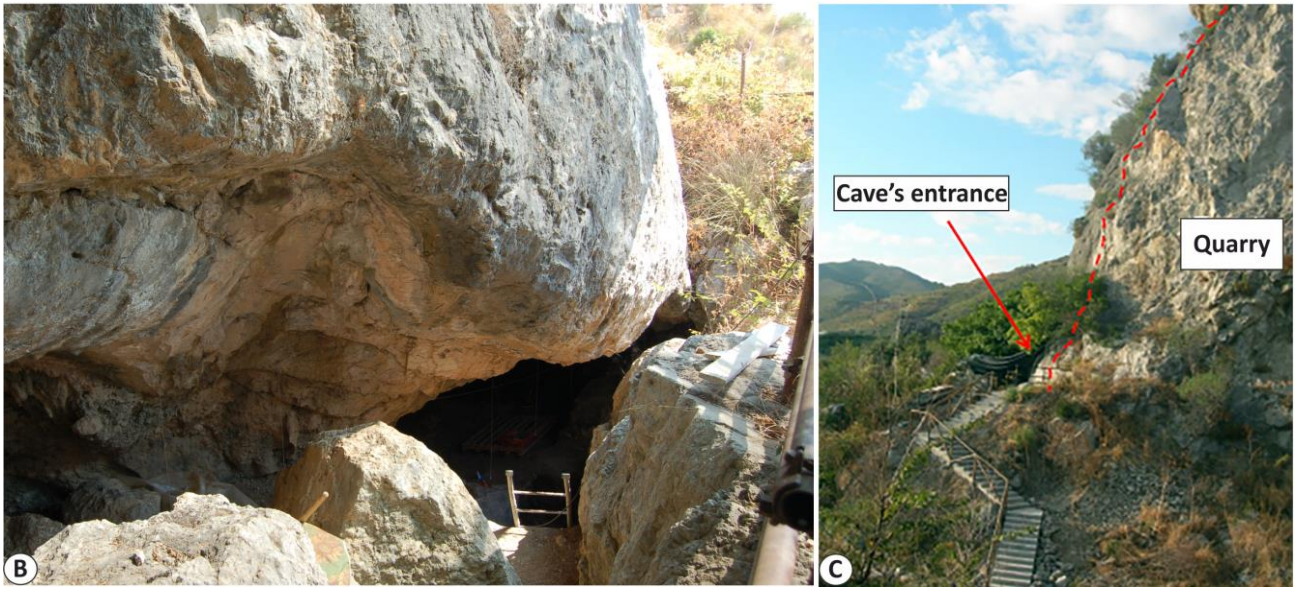

**Figure 3:** A: Location of the Roccia San Sebastiano cave. Note the plain in front of Mt. Massico and the limits of the old quarry (red curve). B: The entrance of the cave. C: Relationship between the quarry front (red dashed line) and the cave entrance (photos by C. Collina).

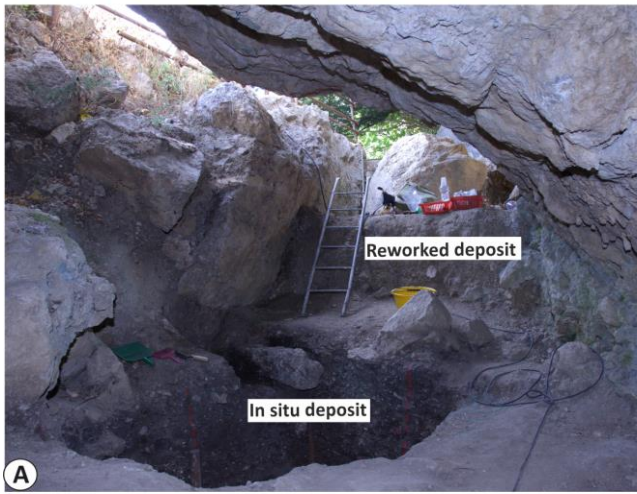
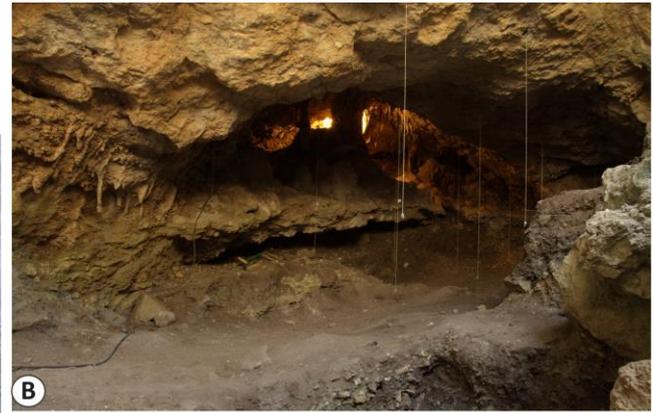
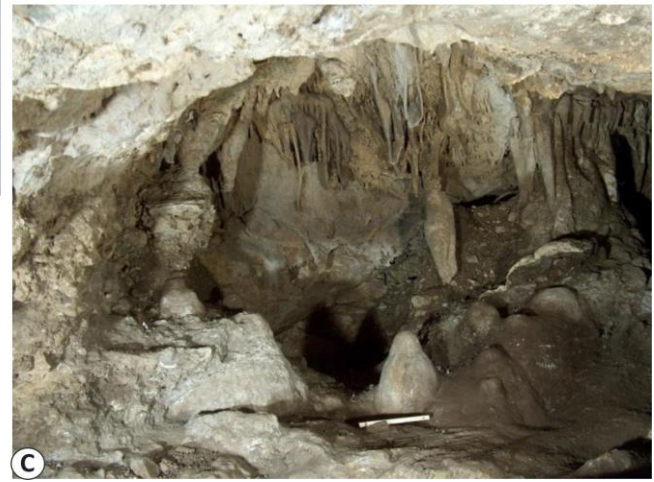

**Figure 4:** A: Panoramic view of the shelter area. See the location of the exacavated archeological trench and the large collapsed block on the left of the stair. B-C. View of the interior cave area (photos by C. Collina). B, sector covered by reworked deposits; C, carbonate stalactites and stalagmites.

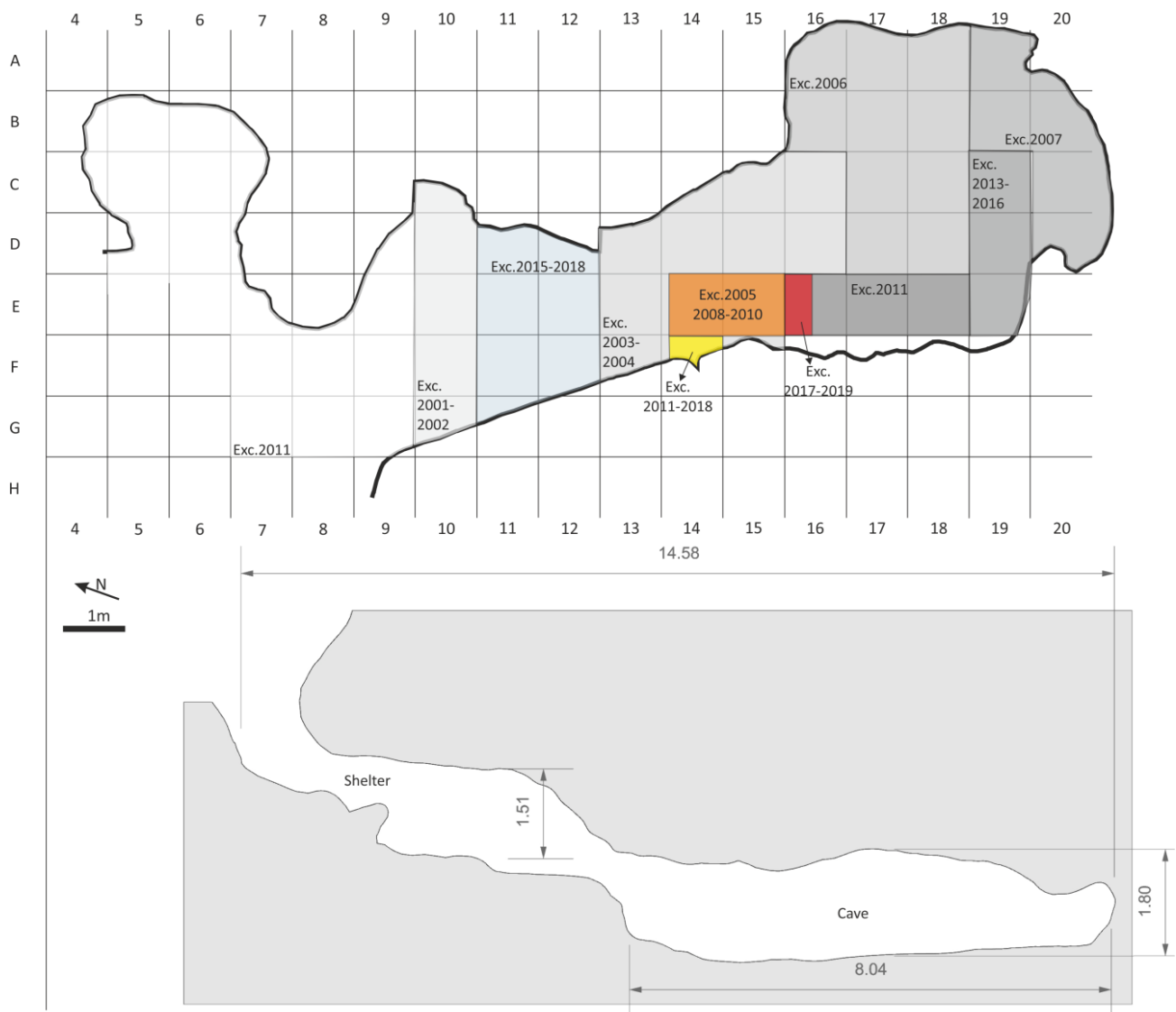

**Figure 5:** Planimetry of the cave (modified after Pennacchioni) showing the excavated areas. Longitudinal section of the cave obtained by ortho-photo of the cave produced by 3D laser scanning (drawing L. Repola).

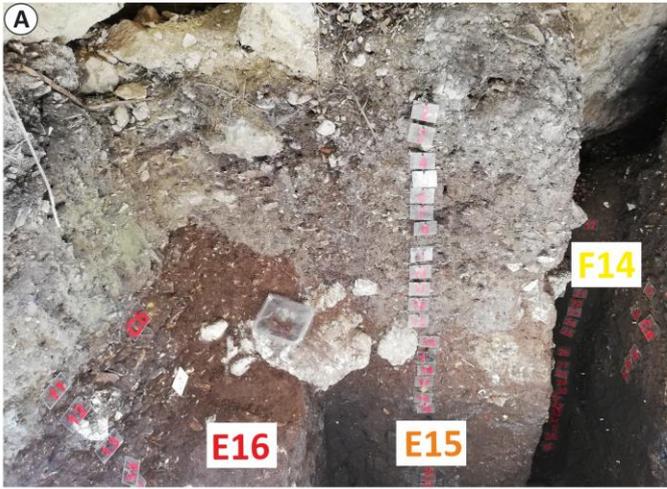
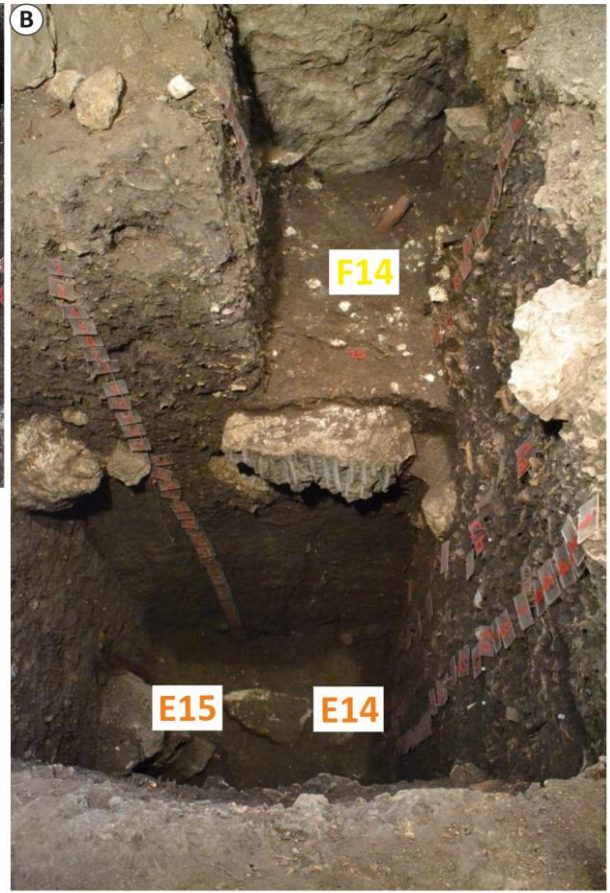
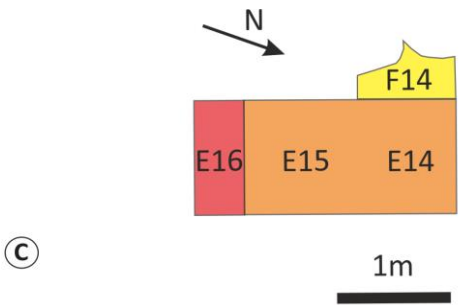
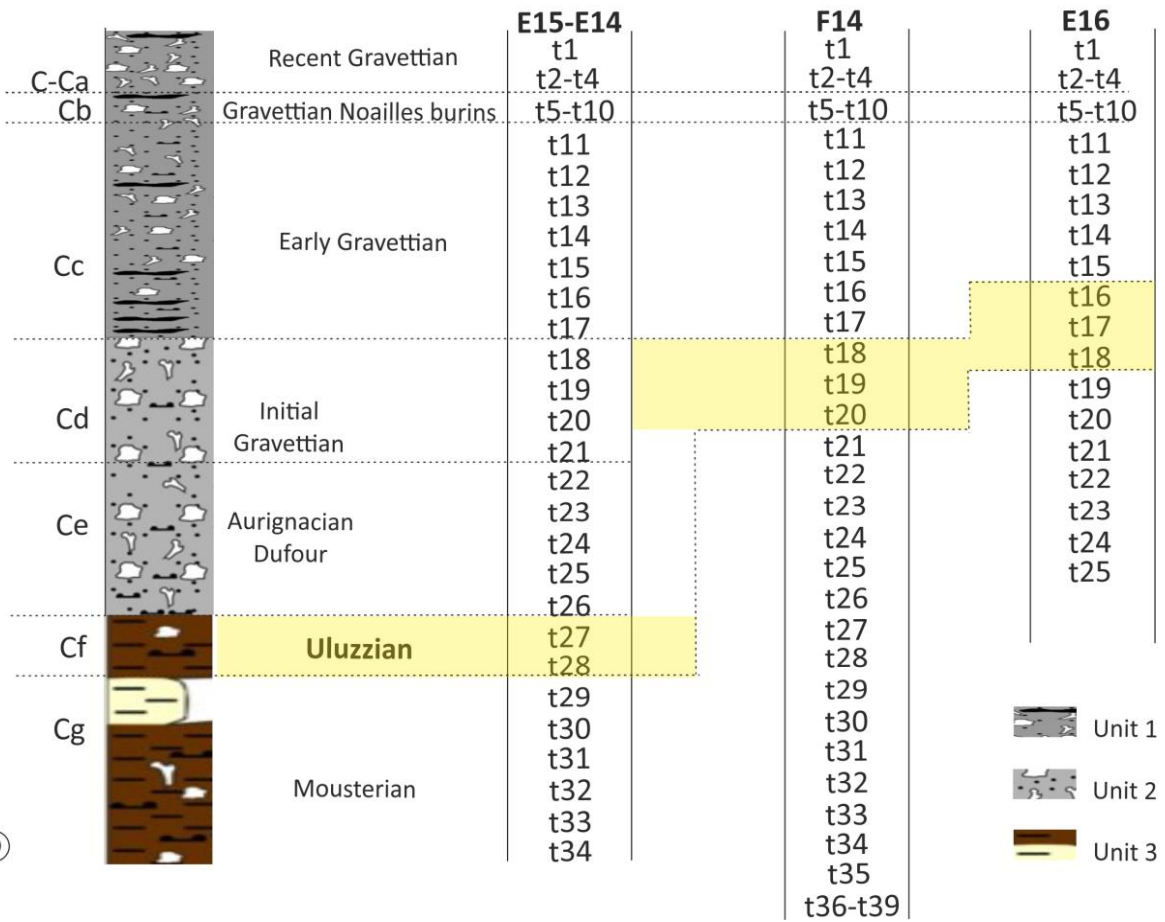

**Figure 6**: A- B: Particular of the excavation area (photos by C. Collina), C: planimetry of the excavated areas, D: stratigraphic sequence correlation, lithological column, cultural domain (highlighted in yellow the Uluzzian levels).

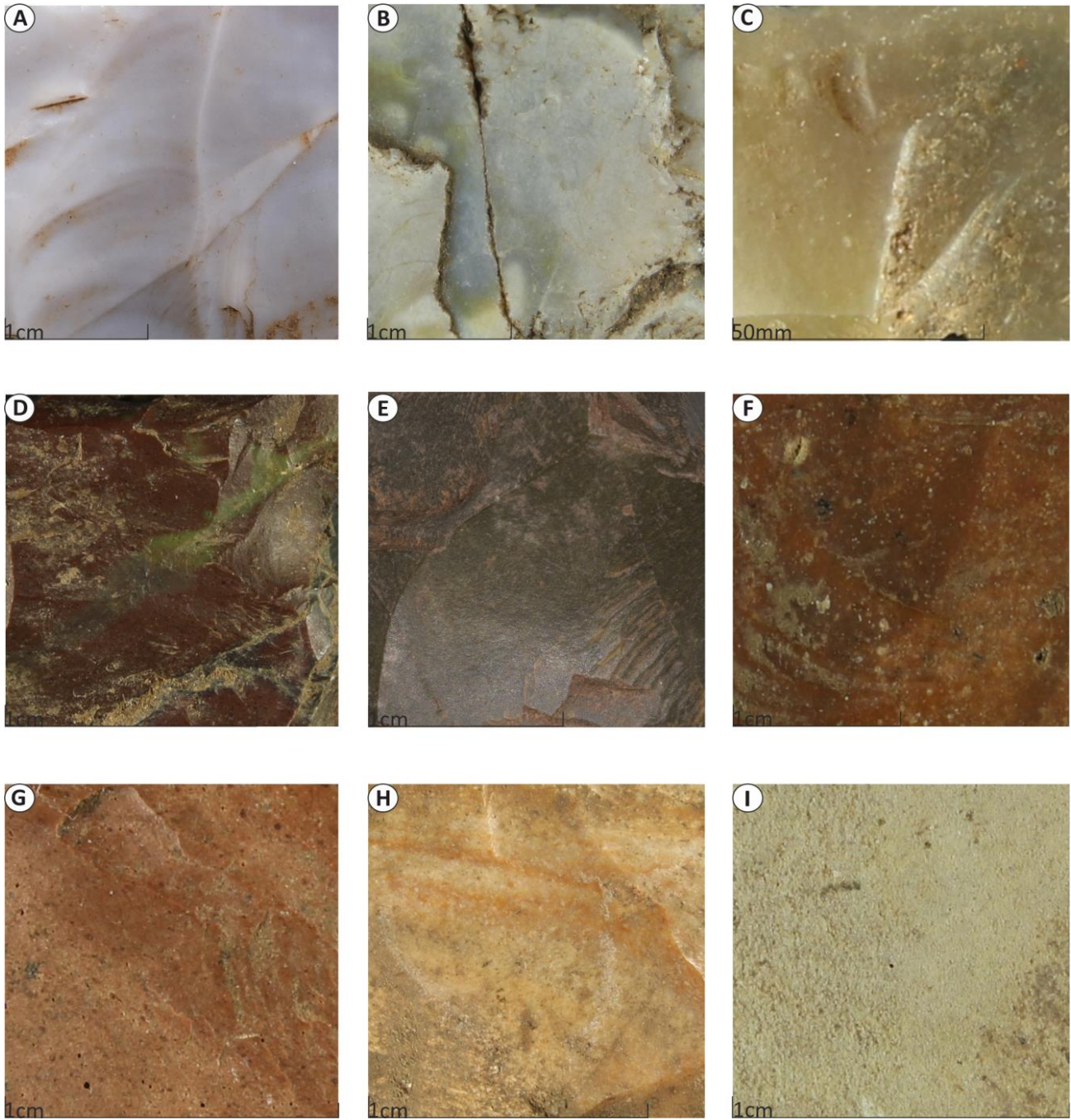

**Figure 7:** Raw material lithotypes. A- B: fine-grained, white/azure, opaque chert; C: fine grained, glossy beige chert; D, E: radiolarite; F: Scaglia Rossa chert; G: cherty limestone; H: quartz-arenite; I: limestone.

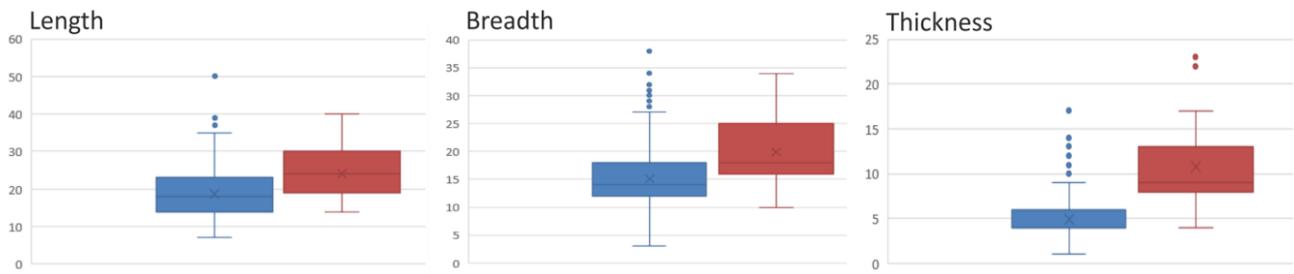

**Figure 8:** Boxplot showing the distribution (in mm) of length, breadth, and thickness in flakes (blue) and cores (red).

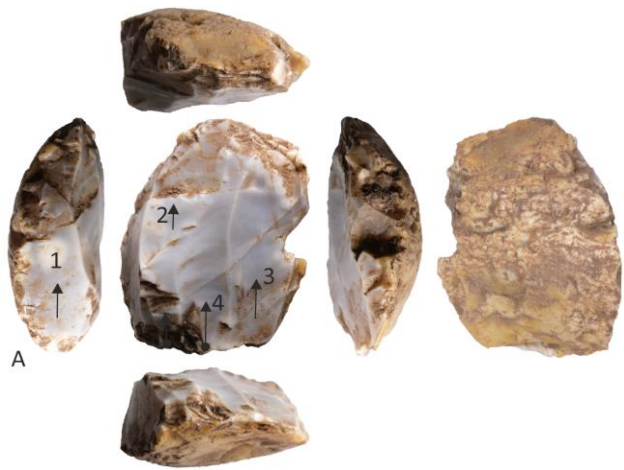

| Selection | Pebble |
|---|---|
| Striking platform | Cortical unprepared<br>Ortoghonal to the debitage surface |
| Production & Target objects | 1, 2, 3, 4 = target objects<br>Medium stage of exploitation<br>Elongated flakes |
| Technique of percussion | Direct freehand percussion<br>Bipolar on anvil |

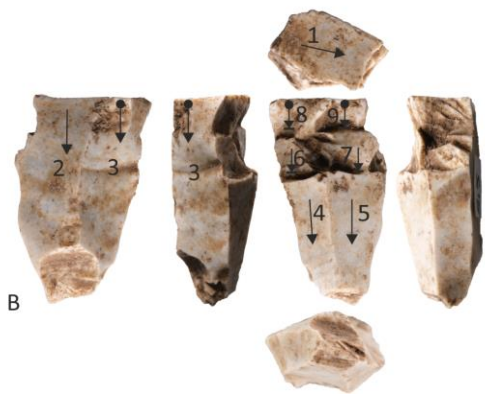

| Selection | Fragment |
|---|---|
| Striking platform | 1 = Made with one stroke<br>Ortoghonal to the debitage surface |
| Production & Target objects | 2, 3, 4 ,5 = target objects<br>6, 7, 8, 9 = hinged accidents<br>Exploited stage<br>Bladelets |
| Technique of percussion | Direct freehand percussion<br>Bipolar on anvil |

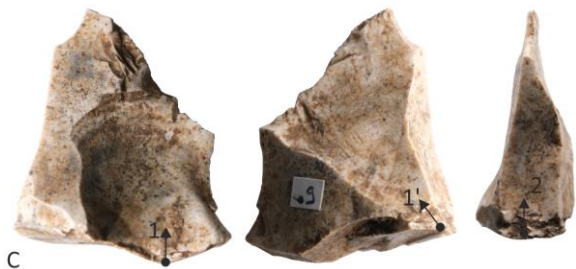

| Selection | Flake |
|---|---|
| Striking platform | Unprepared<br>ortoghonal to the debitage surface |
| Production & Target objects | 1, 1', 2 = Detachements on three side of the flake without any management of convexities or striking platform<br>Exploited stage<br>Flakes and Bladelets |
| Technique of percussion | Direct freehand percussion |

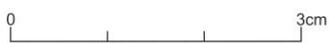 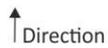 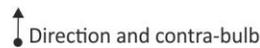 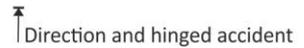

**Figure 9:** Unidirectional core. A: unidirectional core on pebbles, reduction performed by bipolar and freehand percussion technique; B: unidirectional core on fragment, reduction performed by bipolar and freehand percussion technique; C: core on flake.

| Rectilinear longitudinal profile of ventral face | Very similar ventral and dorsal faces |
|---|---|
| 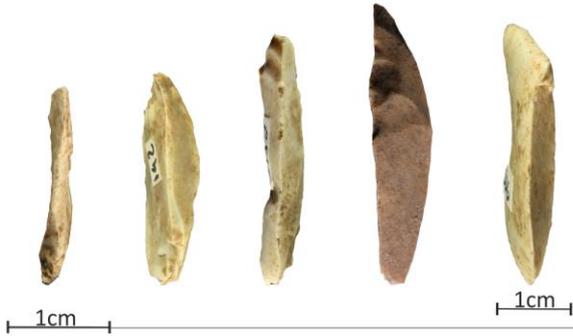 | 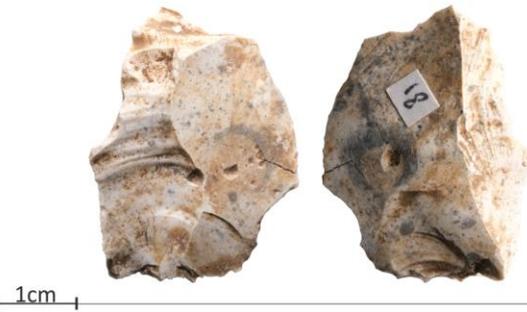 |
| Very pronounced ripple marks | Shattered butts |
| 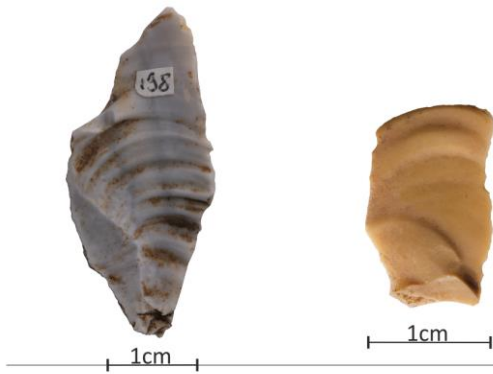 | 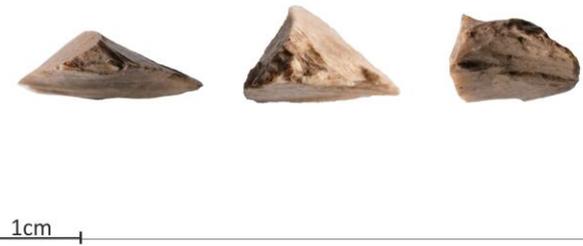 |
| Point-form butts | Linear butts |
| 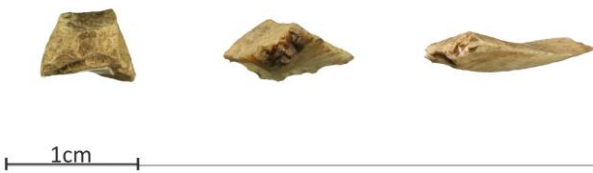 | 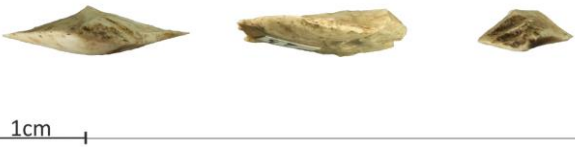 |
| Diffused impact point | Sheared bulbs of percussion / Presence of parasital scar |
| 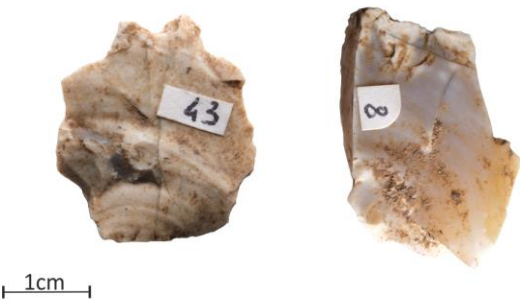 | 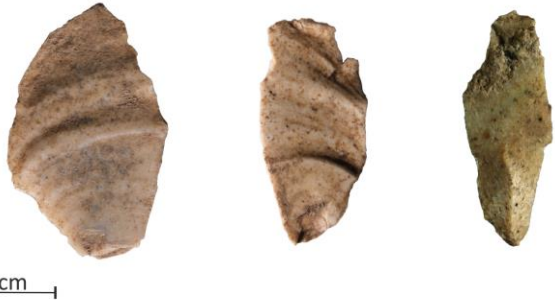 |

**Figure 10:** Main technical criteria for the recognition of bipolar technique of debitage.

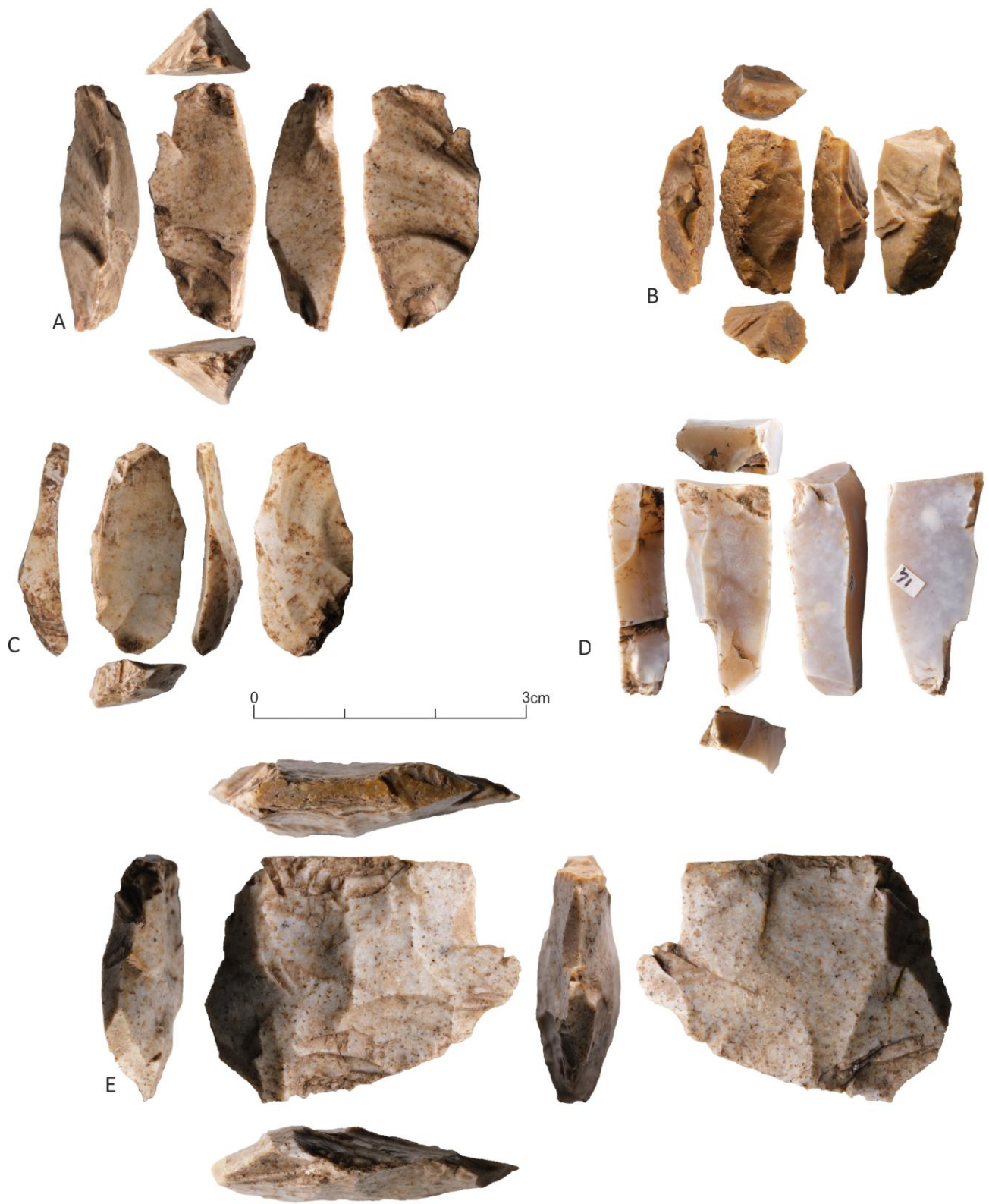

**Figure 11:** Core and products of bipolar technique on anvil. A, B: *bâtonnets*, C: flake; D: core fragment, D: splintered piece (cfr. paragraph 5.1).

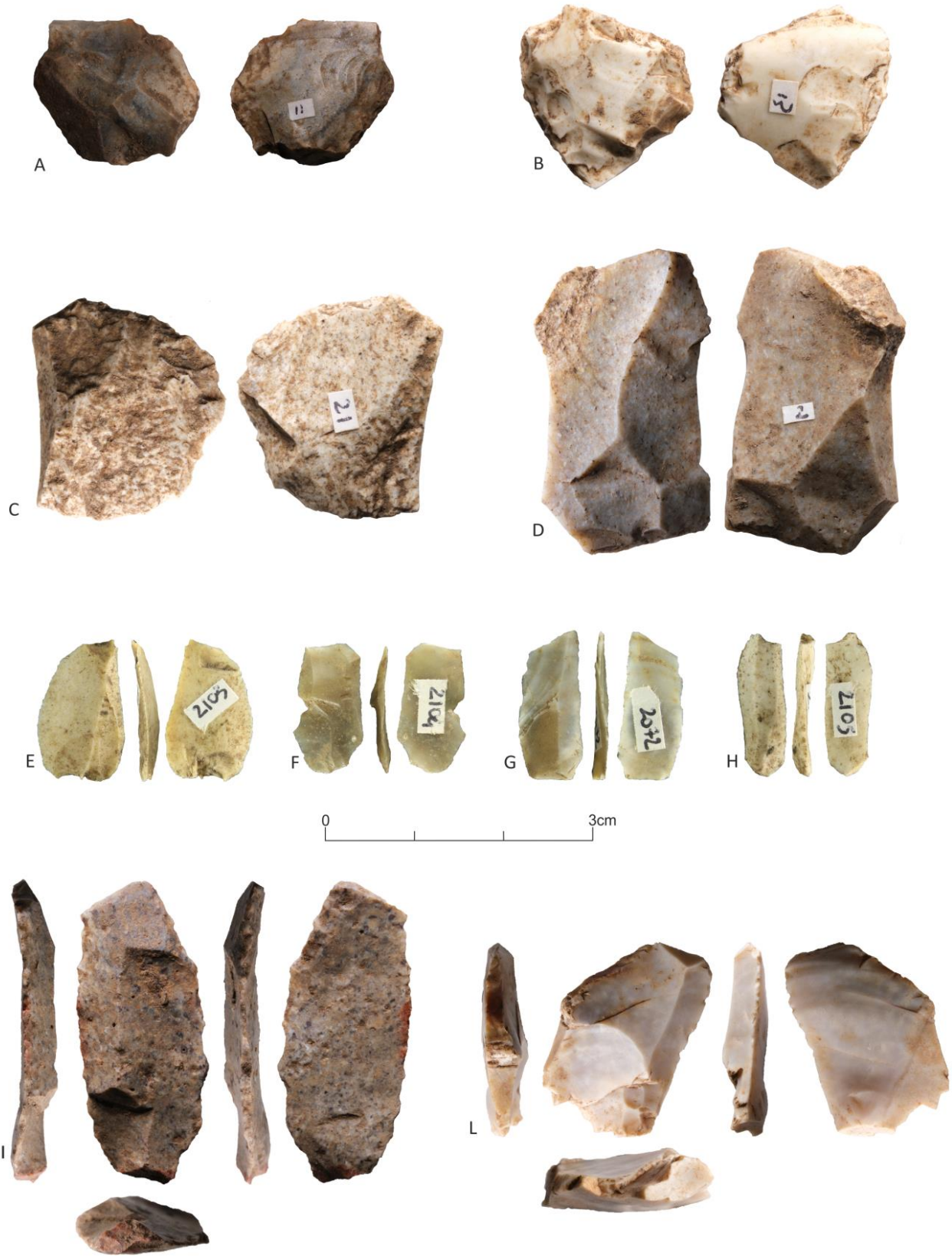

**Figure 12:** Products of debitage: A, B, C, E, L: flakes; D, I: blades; F, G, H: bladelets.

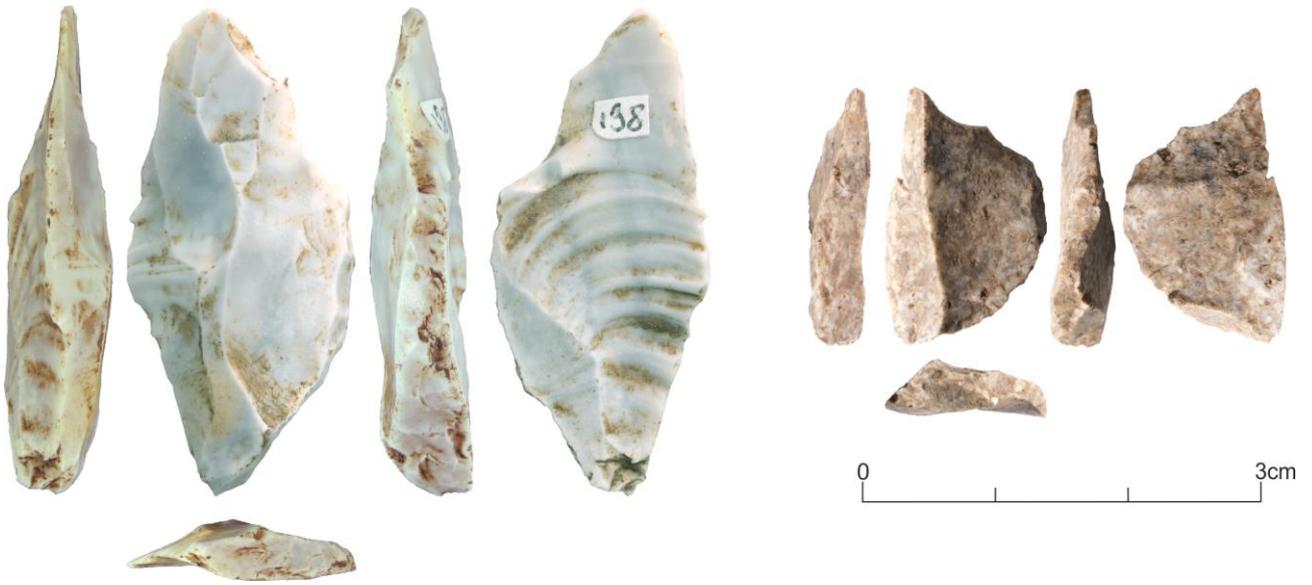

**Figure 13**: The two lunates.

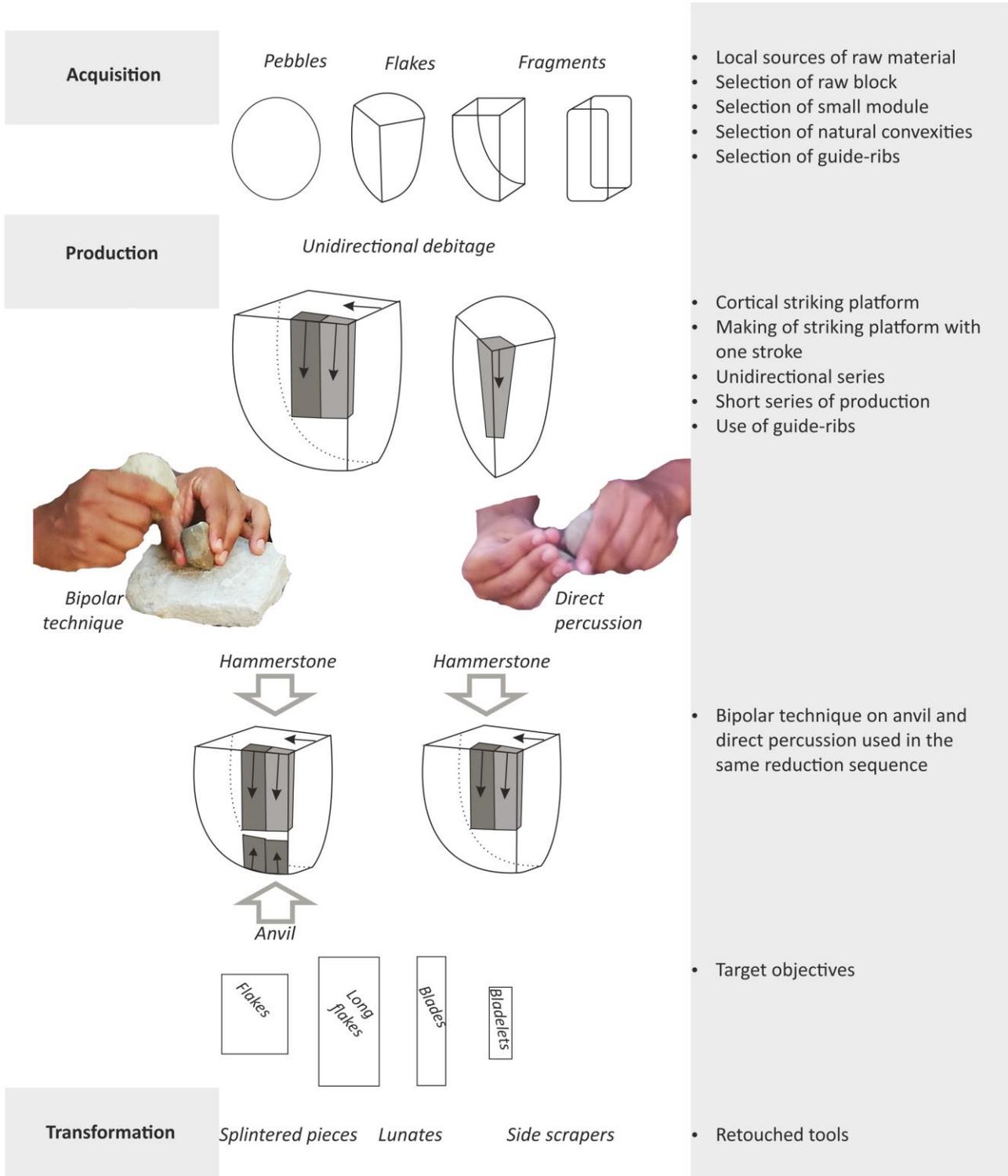

**Figure 14:** Synoptic scheme of the Roccia San Sebastiano Uluzzian reduction sequence.

# Table Captions

| Cultural phase | Sub unit | E 14-E15 | F14 | E16 | Typological description |
|---|---|---|---|---|---|
| Recent Gravettian | Ca | t 1-4 | t 1-4 | t 1-4 | A high percentage of burins, micro-lithic, hyper-micro-lithic points, and backed blades. |
| Gravettian with Noailles burins | Cb | t 5-10 | t 5-10 | t 5-10 | Moderate percentage (1-5%) of Noailles and paranoailles burins. Abundant presence of micro-lithic and hyper-micro-lithic points and backed blades. |
| Early Gravettian | Cc | t 11-17 | t 11-17 | t 11-15 | Decrease in burins and hyper-micro-lithic backed elements with a high percentage of backed tools, reaching about 60% of the assemblage. |
| Initial Gravettian | Cd | t 19-21 | Absent | Absent | Production of marginal backed elements. Absence of hyper-micro-lithic backed elements. Increase in denticulates and notches. Increase of pieces with marginal retouch among the end-scrapers. Bone and antler objects are also present. |
| Aurignacian with Dufour bladelets | Ce | t 22-26 | Absent | Absent | Presence of the typical Dufour bladelets and pieces with marginal retouch with an increase in bone artefacts. Production of bladelets with trapezoidal section obtained from small pebble cores. |
| Uluzzian | Cf | t 27-28 | t 18-20 | t 16-18 | High frequency of splintered pieces obtained by bipolar flaking. Retouched tools include side scrapers, points, backed blades, end scrapers. Presence of lunates with a steep curved backed side. |
| Final Mousterian | Cg | t 29-34 | t 21-39 | t 19-25 | An abundance of points, and side scrapers and notches. The dominant presence of the Levallois debitage, allowing to obtain flakes with strongly predetermined dimensional, technical, and morphological features, whose butt is often facetted or dihedral. |

**Table 1:** Synthetic relationship between the stratigraphic trenches, lithological units and cultural domain (highlighted in yellow the spits that are the object of this study). The typological description is based on the material of the trench E14-E15.

| Cultural phase | Name | Unmodelled (BP) | | | | | | |
|---|---|---|---|---|---|---|---|---|
| | | from | to | % | from | to | % | m |
| Recent Gravettian, E 14-E15, Ca t 1-4 | R_Date Rome-2447 | 23,870 | 23,320 | 68.2 | 24,070 | 23,020 | 95.4 | 23,570 |
| Final Mousterian, E 14-E15, Cg, t 29-34 | R_Date Rome-2111 | 43,680 | 42,190 | 68.2 | 44,740 | 41,700 | 95.4 | 43,010 |

**Table 2:** date from Aiello et al., 2018, calibration Oxcal 4.3, Intcal13 Reimer et al., 2013.

| Trench | Spit | N | % | Total | % |
|---|---|---|---|---|---|
| E16 | t 16 | 208 | 6.4 | 705 | 21.6 |
| | t 17 | 183 | 5.6 | | |
| | t 18 | 314 | 9.6 | | |
| F14 | t 18 | 1153 | 35.4 | 2552 | 78.4 |

|  | t 19 | 992 | 30.5 |  |  |
|---|---|---|---|---|---|
|  | t 20 | 407 | 12.5 |  |  |
| Total | | | 100.0 | **3257** | 100.0 |

**Table 3:** Materials that come from excavation trench E16 spits t 16, t 17, t 18 and from excavation trench F14 spits t 18; t 19; t 20.

| Burned | E16_16 | | E16_17 | | E16_18 | | F14_18 | | F14_19 | | F14_20 | | Total | |
|---|---|---|---|---|---|---|---|---|---|---|---|---|---|---|
|  | n | % | n | % | n | % | n | % | n | % | n | % | n | % |
| No | 146 | 70.2 | 110 | 60.1 | 236 | 75.2 | 1015 | 88.0 | 799 | 80.5 | 363 | 89.2 | 2669 | 81.9 |
| Yes | 62 | 29.8 | 73 | 39.9 | 78 | 24.8 | 138 | 12.0 | 193 | 19.5 | 44 | 10.8 | 588 | 18.1 |
| Total | **208** | 100.0 | **183** | 100.0 | **314** | 100.0 | **1153** | 100.0 | **992** | 100.0 | **407** | 100.0 | **3257** | 100.0 |

**Table 4:** Fire marks.

| Lithotype | E16_16 | | E16_17 | | E16_18 | | F14_18 | | F14_19 | | F14_20 | | Total | |
|---|---|---|---|---|---|---|---|---|---|---|---|---|---|---|
|  | n | % | n | % | n | % | n | % | n | % | n | % | n | % |
| Chert | 199 | 95.7 | 173 | 94.5 | 293 | 93.3 | 1108 | 96.1 | 816 | 82.3 | 344 | 84.5 | **2933** | 90.1 |
| Cherty limestone | 4 | 1.9 | 2 | 1.1 | 3 | 1.0 | 37 | 3.2 | 59 | 5.9 | 23 | 5.7 | **128** | 3.9 |
| Limestone | 1 | 0.5 | 3 | 1.6 | 16 | 5.1 | 2 | 0.2 | 105 | 10.6 | 37 | 9.1 | **164** | 5.0 |
| Quartz-arenite | 2 | 1.0 | 1 | 0.5 | 0 | 0.0 | 1 | 0.1 | 4 | 0.4 | 2 | 0.5 | **10** | 0.3 |
| Radiolarite | 2 | 1.0 | 4 | 2.2 | 2 | 0.6 | 1 | 0 | 8 | 0.8 | 1 | 0.2 | **18** | 0.6 |
| Quartz | 0 | 0.0 | 0 | 0.0 | 0 | 0.0 | 2 | 0.2 | 0 | 0.0 | 0 | 0.0 | **2** | 0.1 |
| Sandstone | 0 | 0.0 | 0 | 0.0 | 0 | 0.0 | 2 | 0.2 | 0 | 0.0 | 0 | 0.0 | **2** | 0.1 |
| Total | **208** | 100.0 | **183** | 100.0 | **314** | 100.0 | **1153** | 100.0 | **992** | 100.0 | **407** | 100.0 | **3257** | 100.0 |

**Table 5:** Raw material lithotypes.

| Granulometry | E16_16 | | E16_17 | | E16_18 | | F14_18 | | F14_19 | | F14_20 | | Total | |
|---|---|---|---|---|---|---|---|---|---|---|---|---|---|---|
|  | n | % | n | % | n | % | n | % | n | % | n | % | n | % |
| Coarse | 9 | 9.4 | 14 | 12.8 | 19 | 10.7 | 40 | 11.1 | 108 | 25.3 | 29 | 18.7 | 219 | 16.5 |
| Fine | 87 | 90.6 | 95 | 87.2 | 159 | 89.3 | 320 | 88.9 | 319 | 74.7 | 126 | 81.3 | 1106 | 83.5 |
| Total | **96** | 100.0 | **109** | 100.0 | **178** | 100.0 | **360** | 100.0 | **427** | 100.0 | **155** | 100.0 | **1325** | 100.0 |

**Table 6:** Raw material granulometry.

| Integrity | E16_16 | | E16_17 | | E16_18 | | F14_18 | | F14_19 | | F14_20 | | Total | |
|---|---|---|---|---|---|---|---|---|---|---|---|---|---|---|
|  | n | % | n | % | n | % | n | % | n | % | n | % | n | % |
| Integer | 69 | 33.2 | 56 | 30.6 | 108 | 34.4 | 299 | 25.9 | 331 | 33.4 | 158 | 38.8 | **1021** | 31.3 |
| Composite | 104 | 50.0 | 83 | 45.4 | 138 | 43.9 | 766 | 66.4 | 545 | 54.9 | 201 | 49.4 | **1837** | 56.4 |
| Distal | 18 | 8.7 | 18 | 9.8 | 19 | 6.1 | 28 | 2.4 | 37 | 3.7 | 18 | 4.4 | **138** | 4.2 |
| Lateral | 1 | 0.5 | 2 | 1.1 | 1 | 0.3 | 9 | 0.8 | 11 | 1.1 | 0 | 0.0 | **24** | 0.7 |

| | | | | | | | | | | | | | |
|---|---|---|---|---|---|---|---|---|---|---|---|---|---|
| Mesial | 6 | 2.9 | 8 | 4.4 | 14 | 4.5 | 17 | 1.5 | 15 | 1.5 | 6 | 1.5 | **66** | 2.0 |
| Proximal | 10 | 4.8 | 16 | 8.7 | 34 | 10.8 | 34 | 2.9 | 53 | 5.3 | 24 | 5.9 | **171** | 5.3 |
| Total | **208** | 100.0 | **183** | 100.0 | **314** | 100.0 | **1153** | 100.0 | **992** | 100.0 | **407** | 100.0 | **3257** | 100.0 |

**Table 7:** Integrity of the items and location of the fracture.

| DC | E16_16 | | E16_17 | | E16_18 | | F14_18 | | F14_19 | | F14_20 | | Total | |
|---|---|---|---|---|---|---|---|---|---|---|---|---|---|---|
| | n | % | n | % | n | % | n | % | n | % | n | % | n | % |
| 1 | 41 | 19.7 | 19 | 10.4 | 43 | 13.7 | 489 | 42.4 | 208 | 21.0 | 86 | 21.1 | 886 | 27.2 |
| 2 | 82 | 39.4 | 49 | 26.8 | 90 | 28.7 | 318 | 27.6 | 259 | 26.1 | 137 | 33.7 | 935 | 28.7 |
| 3 | 21 | 10.1 | 41 | 22.4 | 59 | 18.8 | 123 | 10.7 | 196 | 19.8 | 98 | 24.1 | 538 | 16.5 |
| 4 | 20 | 9.6 | 20 | 10.9 | 37 | 11.8 | 75 | 6.5 | 96 | 9.7 | 38 | 9.3 | 286 | 8.8 |
| 5 | 44 | 21.2 | 54 | 29.5 | 85 | 27.1 | 148 | 12.8 | 233 | 23.5 | 48 | 11.8 | 612 | 18.8 |
| Total | **208** | 100.0 | **183** | 100.0 | **314** | 100.0 | **1153** | 100.0 | **992** | 100.0 | **407** | 100.0 | **3257** | 100.0 |

**Table 8:** Dimensional classes - DC (first: 1-50 mm2, second: 50-100 mm2, third: 100-150 mm2, fourth: 150-200 mm2, fifth: > 200 mm2).

| Technological Class | E16_16 | | E16_17 | | E16_18 | | F14_18 | | F14_19 | | F14_20 | | Total | |
|---|---|---|---|---|---|---|---|---|---|---|---|---|---|---|
| | n | % | n | % | n | % | n | % | n | % | n | % | n | % |
| **Core** | 3 | 1.4 | 5 | 2.7 | 5 | 1.6 | 14 | 1.2 | 17 | 1.7 | 3 | 0.7 | 47 | 1.4 |
| **Flake** | 76 | 36.5 | 90 | 49.2 | 145 | 46.2 | 224 | 19.4 | 343 | 34.6 | 126 | 31.0 | 1004 | 30.8 |
| **Microflake** | 27 | 13.0 | 19 | 10.4 | 39 | 12.4 | 180 | 15.6 | 123 | 12.4 | 87 | 21.4 | 475 | 14.6 |
| **Debris** | 102 | 49.0 | 69 | 37.7 | 125 | 39.8 | 734 | 63.7 | 509 | 51.3 | 191 | 46.9 | 1730 | 53.1 |
| **Hammerstone** | 0 | 0.0 | 0 | 0.0 | 0 | 0.0 | 1 | 0.1 | 0 | 0.0 | 0 | 0.0 | 1 | 0.0 |
| **Total** | **208** | 100.0 | **183** | 100.0 | **314** | 100.0 | **1153** | 100.0 | **992** | 100.0 | **407** | 100.0 | **3257** | 100.0 |

**Table 9:** Technological classes.

| Raw block | N. | % |
|---|---|---|
| **Flake** | 26 | 55.3 |
| **Fragment** | 14 | 29.8 |
| **Pebble** | 5 | 10.6 |
| **Indeterminate** | 2 | 4.3 |
| **Total** | **47** | 100.0 |

**Table 10:** Raw block.

| Bipolar Technique | | |
|---|---|---|
| **Technological class** | N | % |
| **Core** | 26 | 7.2 |
| **Flake** | 201 | 55.5 |
| **Micro-flake** | 34 | 9.4 |
| **Debris** | 101 | 27.9 |
| **Total** | **362** | 100.0 |

**Table 11:** Technological classes showing the use of bipolar percussion technique.

| Longitudinal profile | | | Scar direction | | | Butt | | | Impact point | | | Bulb | | |
|---|---|---|---|---|---|---|---|---|---|---|---|---|---|---|
| Trait | N | % | Trait | N | % | Trait | N | % | Trait | N | % | Trait | N | % |
| Concave | 4 | 2.0 | Cortical | 2 | 1.0 | Cortical | 17 | 8.5 | Diffuse | 79 | 39.3 | Flat | 141 | 70.1 |
| Convex | 6 | 3.0 | Bidirectional | 72 | 35.8 | Linear | 69 | 34.3 | Central | 31 | 15.4 | Prominent | 11 | 5.5 |
| Rectilinear | 181 | 90.0 | Unidirectional | 77 | 38.3 | Point form | 26 | 12.9 | Lateral | 37 | 18.4 | Double, other | 4 | 2.0 |
| Twisted | 3 | 1.5 | Ventral face | 19 | 9.5 | Flat | 23 | 11.4 | Broken | 34 | 16.9 | Broken | 35 | 17.4 |
| Wavy | 7 | 3.5 | Orthogonal | 5 | 2.5 | Sheared | 21 | 10.4 | Indeterminate | 20 | 10.0 | Indeterminate | 10 | 5.0 |
| | | | Perpendicular | 3 | 1.5 | Prepared-dihedral | 4 | 2.0 | | | | | | |
| | | | Convergent | 1 | 0.5 | Broken | 30 | 14.9 | | | | | | |
| | | | Indeterminate | 22 | 10.9 | Indeterminate | 11 | 5.5 | | | | | | |
| **Total** | **201** | **100.0** | | **201** | **100.0** | | **201** | **100.0** | | **201** | **100.0** | | **201** | **100.0** |

**Table 12:** Technical traits of flakes obtained by bipolar percussion technique.

| Technological categories | N | % |
|---|---|---|
| Completely cortical flake | 16 | 1.6 |
| Semi cortical flake | 46 | 4.6 |
| Completely cortical fragmented flake | 19 | 1.9 |
| Semi cortical fragmented flake | 58 | 5.8 |
| Blade | 41 | 4.1 |
| Long flake | 69 | 6.9 |
| Flake | 258 | 25.7 |
| Composite flake | 104 | 10.4 |
| Distal flake | 94 | 9.4 |
| Lateral flake | 19 | 1.9 |
| Mesial flake | 52 | 5.2 |
| Proximal flake | 146 | 14.5 |
| Debordant flake, pseudo Levallois point | 19 | 1.9 |
| Fragmented debordant flake | 3 | 0.3 |
| *Bâtonnets* | 23 | 2.3 |
| Indeterminate | 37 | 3.7 |
| **Integer** | **486** | **48.4** |
| **Fragmented** | **518** | **51.6** |
| **Total** | **1004** | **100,0** |

**Table 13:** Technological categories of flakes.

| Direction | N | % |
|---|---|---|

| | N | % |
|---|---|---|
| Unidirectional | 541 | 53.9 |
| Bidirectional | 92 | 9.2 |
| Orthogonal | 54 | 5.4 |
| Convergent | 36 | 3.6 |
| Cortical | 38 | 3.8 |
| Perpendicular | 16 | 1.6 |
| Centripetal | 4 | 0.4 |
| Ventral face | 46 | 4.6 |
| Indeterminate | 177 | 17.6 |
| Total | 1004 | 100.0 |

Table 14: Directions of removals.

| Butt | N | % |
|---|---|---|
| Flat | 307 | 30.6 |
| Linear | 146 | 14.5 |
| Cortical | 52 | 5.2 |
| Dihedral | 43 | 4.3 |
| Point form | 43 | 4.3 |
| Prepared | 37 | 3.7 |
| Broken | 26 | 2.6 |
| Facetted | 16 | 1.6 |
| Indeterminate | 48 | 4.8 |
| Absent | 286 | 28.5 |
| Total | 1004 | 100.0 |

Table 15: Types of butt.

| Retouched items\Technological class | Core | Debris | Flake | Total | % |
|---|---|---|---|---|---|
| Splintered pieces | | 5 | 19 | 24 | 19.7 |
| Side scrapers | 3 | 8 | 18 | 29 | 23.8 |
| End scrapers | | 3 | 5 | 8 | 6.6 |
| Points | | 1 | 7 | 8 | 6.6 |
| Backed items | | 1 | 6 | 7 | 5.7 |
| Lunates | | | 2 | 2 | 1.6 |
| Biside scrapers | | | 4 | 4 | 3.3 |
| Denticulates | | | 2 | 2 | 1.6 |
| Notches | | 1 | 4 | 5 | 4.1 |
| Transversal scrapers | | | 2 | 2 | 1.6 |
| Yes | 1 | 22 | 8 | 31 | 25.4 |
| Total | 4 | 41 | 77 | 122 | 100.0 |
| % | 3.3 | 33.6 | 63.1 | 100.0 | |

Table 16: Typology of retouched tools.

| Main characteristics of the Uluzzian techno-complex | Roccia San Sebastiano | Cavallo | Uluzzo C | Castelcivita | Colleroton | Fabbrica | Broion | Fumane |
|---|---|---|---|---|---|---|---|---|
| Local raw materials | **Yes** | Y | Y | Y | Y | Y | Y | Y |
| Additional/simple concepts of debitage (unidirectional and orthogonal methods) | **Yes** | Y | Y | Y | Y | Y | Y | Y |
| Absence of integrated concepts (Levallois, blade/bladelets reduction systems UP) | **Yes** | Y | N | Y | Y | Y | Y | N |
| Dominated use of the bipolar technique on anvil | **Yes** | Y | Y | Y | Y | Y | Y | N |
| Production of several morphologies of flakes and elongated pieces | **Yes** | Y | Y | Y | Y | Y | Y | Y |
| Low degree of standardisation of the products | **Yes** | Y | Y | Y | Y | Y | Y | Y |
| Lunates, curved backed pieces | **Yes** | Y | Y | Y | Y | Y | Y | N |
| Systematic production of end-scrapers | **No** | Y | Y | Y | Y | Y | Y | Y |

**Table 17:** Main characteristics of the Uluzzian techno-complex (Moroni et al., 2018; Marciani et al., 2020). Y = yes; N= no. Cavallo (Palma di Cesnola, 1963; 1964;1966; Moroni et al., 2018); Uluzzo C (unpublished); Castelcivita, (Gambassini, 1997); Colle Rotondo (Villa et al., 2018); Fabbrica (Dini and Tozzi, 2012; Villa et al., 2018); Broion riparo (Peresani et al., 2019); Fumane (Peresani et al., 2016; 2019).